%% file: HFatLHC.tex
\documentclass{elsarticle}

\usepackage{lineno,hyperref}
\usepackage{xspace}
\usepackage{graphicx}
\usepackage{mciteplus}

\modulolinenumbers[5]

\usepackage{ifthen} 
\newboolean{uprightparticles}
\setboolean{uprightparticles}{false} 

\newboolean{inbibliography}
\setboolean{inbibliography}{false} 
\include{lhcb-symbols-def}

\journal{Comptes Rendus de Physique de l'Academie des Sciences}

\biboptions{sort&compress}

\begin{document}

\begin{frontmatter}

\title{Heavy Flavour Physics at the LHC}

\author[Warwick]{T.~Gershon}
\ead{T.J.Gershon@warwick.ac.uk}
\address[Warwick]{Department of Physics, University of Warwick, Coventry, United Kingdom}

\author[Edinburgh]{M.~Needham}
\ead{Matthew.Needham@cern.ch}
\address[Edinburgh]{School of Physics and Astronomy, University of Edinburgh, Edinburgh, United Kingdom}

\begin{abstract}
  A summary of results in heavy flavour physics from Run 1 of the LHC is presented.
  Topics discussed include spectroscopy, mixing, \CP violation and rare decays of charmed and beauty hadrons.
\\ \vspace{-1.0ex}

\noindent
Un r\'{e}sum\'{e} des r\'{e}sultats du run 1 du LHC sur la physique des saveurs lourdes est pr\'{e}sent\'{e}.
Les sujets discut\'{e}s incluent la spectroscopie, le m\'{e}lange entre m\'{e}sons et anti-m\'{e}sons, la violation de \CP et les d\'{e}sint\'{e}grations rares des hadrons charm\'{e}s et de beaut\'{e}. 
\end{abstract}

\begin{keyword}
quark flavour physics
\sep
spectroscopy
\sep
\CP violation 
\sep 
rare decays
\sep 
Large Hadron Collider
\\ \vspace{0.5ex}

\noindent
{\it Mots-cl\'{e}s:} physique des saveurs lourdes, spectroscopie, violation de \CP, d\'{e}sint\'{e}grations rares, Grand collisionneur de hadrons (LHC). 

\end{keyword}

\end{frontmatter}


\section{Introduction}
\label{sec:introduction}

Within the Standard Model of particle physics, there are six ``flavours'' of quark.  
Hadrons containing charm ($c$) or beauty ($b$) quarks or antiquarks are known as heavy flavoured particles.
By a quirk of convention, the heaviest flavoured particle, the top quark, is not usually included with discussions of ``heavy flavour physics'', and thus it is not within the scope of this review.

The high energy proton-proton collisions at the Large Hadron Collider are the world's most copious source of heavy flavoured particles.
For example, the production cross-sections for $c\bar{c}$ and $b\bar{b}$ quark-antiquark pairs are ${\cal O}(1\mbarn)$ and ${\cal O}(100\mub)$ respectively.
Specific values depend on the kinematic region in which the measurement is made, with the production peaking at pseudorapidity values $|\eta| \sim 3$, and are an increasing function of centre-of-mass energy.
Therefore, per $\invfb$ of data collected at, for example, LHCb, over $10^{11}$ $b\bar{b}$ quark pairs are, in principle, available, with the corresponding number for charm pairs an order of magnitude larger.
The number that is actually recorded by each experiment depends critically on the trigger used for online selection.

The availability of such large samples enables a new era of precision flavour physics and discovery.
Measurements of the properties of heavy flavoured hadrons -- including those of states that had not previously been observed -- provide tests of the theory of the strong interaction, quantum chromodynamics (QCD), in a regime where it is not well understood.
In addition, heavy flavour phenomena such as mixing and rare decays are mediated by loop processes involving virtual particles, and are sensitive to possible non-Standard Model physics at high scales.
In fact, precision measurements in the quark sector can be sensitive to new particles with masses much higher than the reach of the LHC collisions. 
Studies of these processes therefore provide a complementary approach to discover ``new physics'', and form a key part of the LHC programme to search beyond the Standard Model.

One particularly enticing aspect of this programme is the possibility to learn more about the origin of the asymmetry between matter and antimatter in the Universe.
Violation of the combined \CP symmetry, that is symmetry under inversion of parity ($P$) and of all internal quantum numbers (charge conjugation, $C$) is a prerequisite for the evolution of an asymmetric Universe.
Within the Standard Model, \CP violation arises due to a complex phase in the Cabibbo-Kobayashi-Maskawa (CKM) quark mixing matrix~\cite{Cabibbo:1963yz,Kobayashi:1973fv} that describes the relative coupling strengths of the charged-current weak interaction transitions between different flavours of quarks.
While this gives a consistent description of all the \CP violation effects that have been observed to date, additional sources of asymmetry must exist to explain the observed Universe.
One of the highest priorities in contemporary particle physics is to discover new sources of \CP violation, which may be present in the quark sector or may be manifest in other areas, for example neutrino oscillations.

\section{Detectors}
\label{sec:detectors}

All the main LHC detectors have considerable potential to study heavy flavour physics, and ATLAS~\cite{Aad:2008zzm}, CMS~\cite{Chatrchyan:2008aa} and LHCb~\cite{Alves:2008zz} have extensive programmes in this area.
Although ALICE can study production at low luminosity, it cannot perform competitive studies of the processes that are the main focus of this review, and therefore is not discussed further.

\begin{figure}[!t]
\centering
\includegraphics[bb=0 0 565 540,clip=true,width=0.27\textwidth]{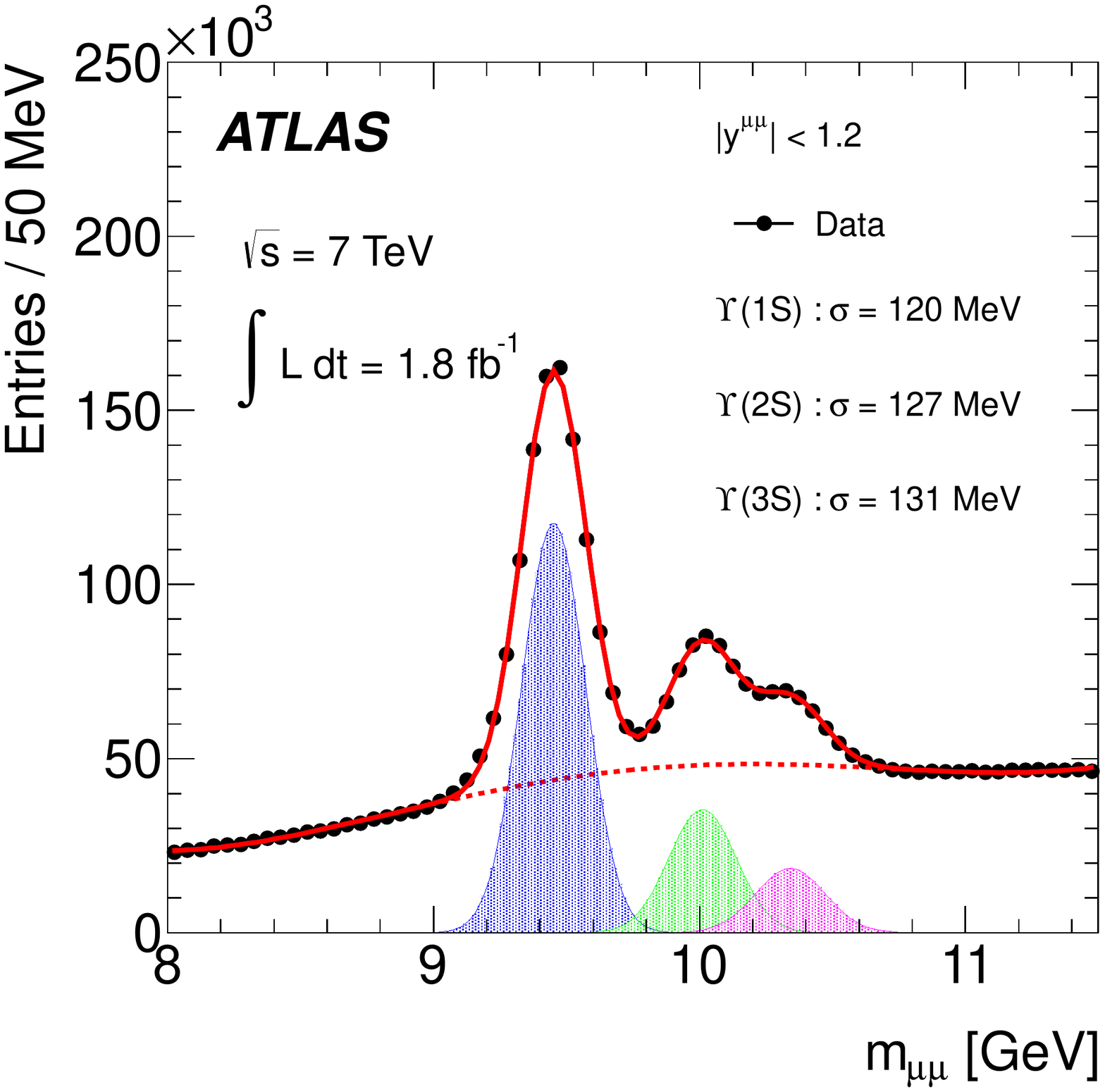}
\includegraphics[width=0.37\textwidth]{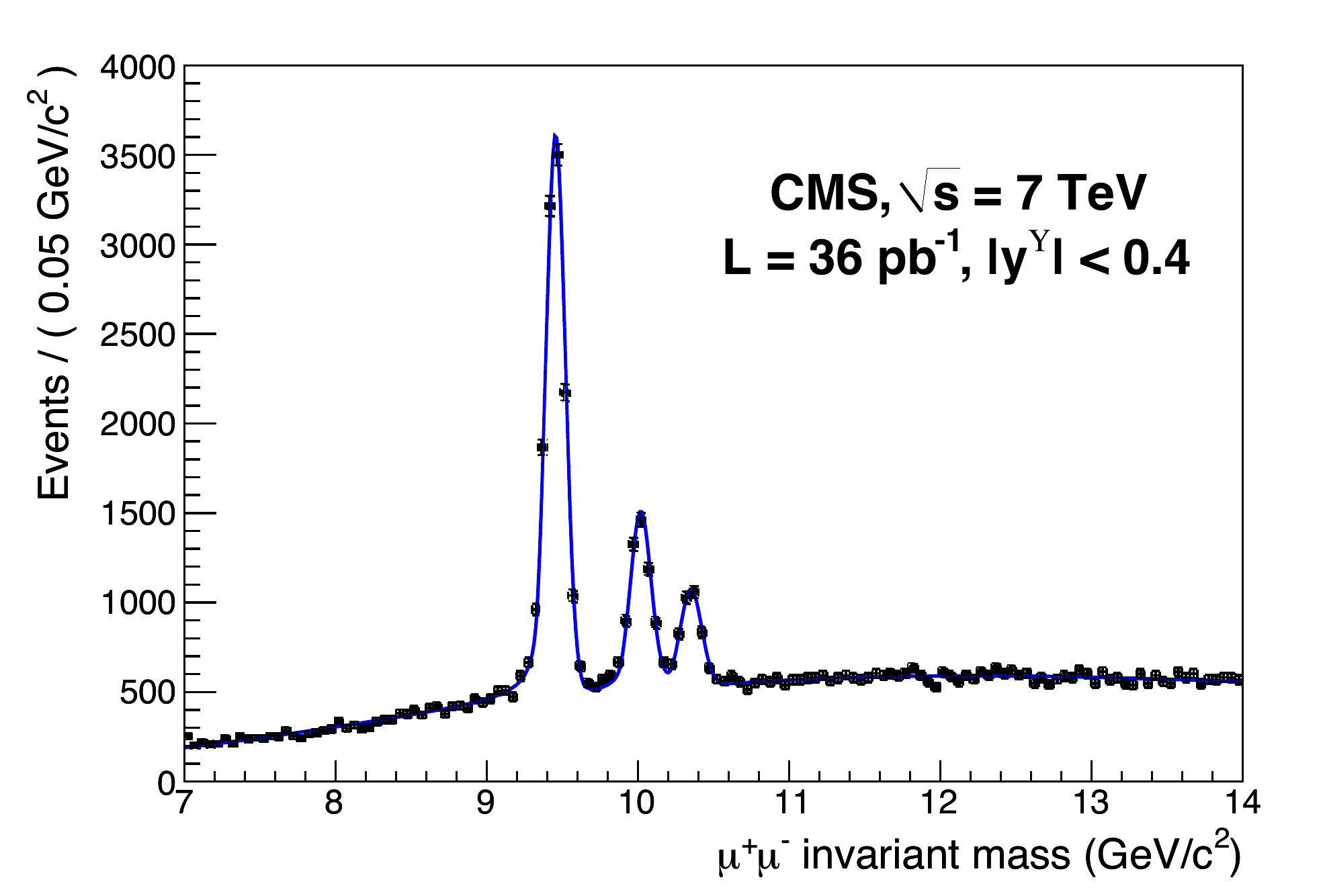}
\includegraphics[width=0.34\textwidth]{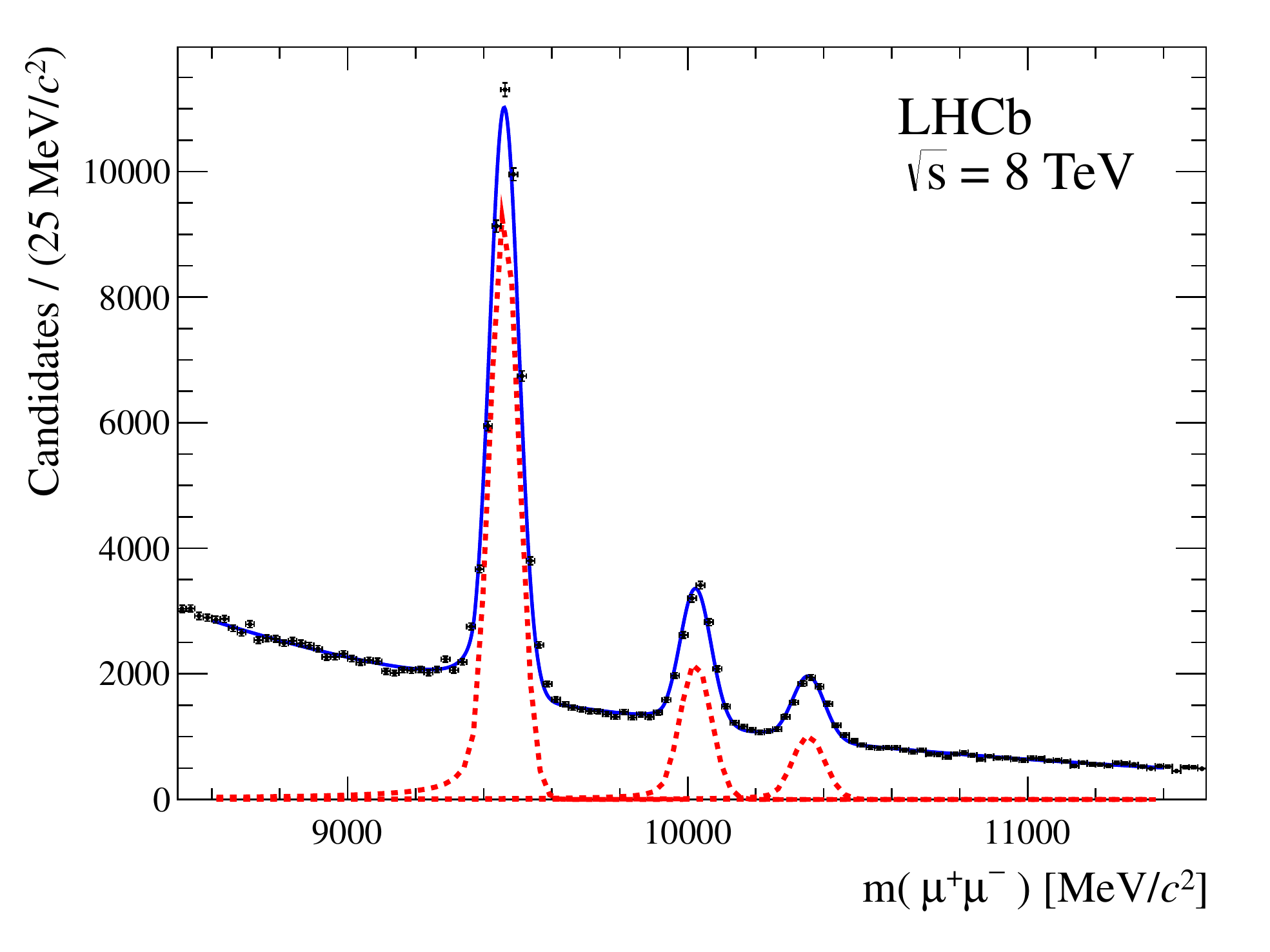}
\caption{\small 
  Dimuon invariant mass resolution in the region of the $\Upsilon$ resonances for (left) ATLAS~\cite{Aad:2012dlq} (middle) CMS~\cite{Chatrchyan:2013yna} (right) LHCb~\cite{LHCb-PAPER-2013-016}. 
}
\label{fig:ups}
\end{figure}

Precision tracking is a key requirement in a hadronic environment in order to reduce combinatorial background to an acceptable level. 
All the LHC detectors have well aligned and calibrated tracking detectors delivering performance close to design expectations. 
An important performance indicator of the tracking system is the mass resolution for the $\Upsilon$ resonances decaying to the dimuon final state, shown in Fig.~\ref{fig:ups}. 
ATLAS, CMS and LHCb have $\Upsilon(1S)$ mass resolutions of $\sim 120 \mevcc$, $\sim 70 \mevcc$ and $\sim 45 \mevcc$ respectively.

The key signatures that allow decays of heavy flavoured particles to be distinguished from random combinations of tracks are the presence of muons with comparatively large transverse momentum, and a displaced vertex due to the non-negligible lifetimes of the weakly decaying hadrons.  
ATLAS and CMS exploit only the former in their online selections, while LHCb makes extensive use in its trigger~\cite{LHCb-DP-2012-004} of information from its vertex locator~\cite{LHCb-DP-2014-001}.
This, together with the particle identification capability provided by its ring-imaging Cherenkov detectors~\cite{LHCb-DP-2012-003}, gives LHCb a broader heavy flavour physics programme than ATLAS and CMS.
However, ATLAS and CMS benefit from higher integrated luminosities ($\sim 5 \invfb$ of $\sqrt{s} = 7 \tev$ and $\sim 20 \invfb$ of $\sqrt{s} = 8 \tev$ $pp$ collisions collected in 2011 and 2012, respectively, for each experiment) and are therefore highly competitive for final states containing dimuon signatures.
LHCb operates at a lower instantaneous luminosity to avoid saturating its trigger and causing over-occupancy of its subdetectors: the corresponding integrated luminosities are $1 \invfb$ (2011) and $2 \invfb$ (2012).
The LHCb upgrade~\cite{LHCb-TDR-012} will enable higher luminosity operation.

\section{Spectroscopy}
\label{sec:spectroscopy}

The large datasets collected at the LHC have allowed studies of the properties of $\bquark$-hadrons with unprecedented precision. 
This is a wide field of research, in which three particularly interesting areas, discussed below, are exotic spectroscopy, the $B_c^+$ meson and $\bquark$-baryons.

In addition to conventional mesons and baryons, the QCD Lagrangian allows for more exotic possibilities such as tetraquark and molecular states. 
The unexpected discovery of the $X(3872)$ state by the Belle collaboration~\cite{Choi:2003ue} led to a resurgence of interest in exotic spectroscopy and subsequently many new ``XYZ'' states have been claimed. 

\begin{figure}[!t]
\centering
\includegraphics[width=0.48\textwidth]{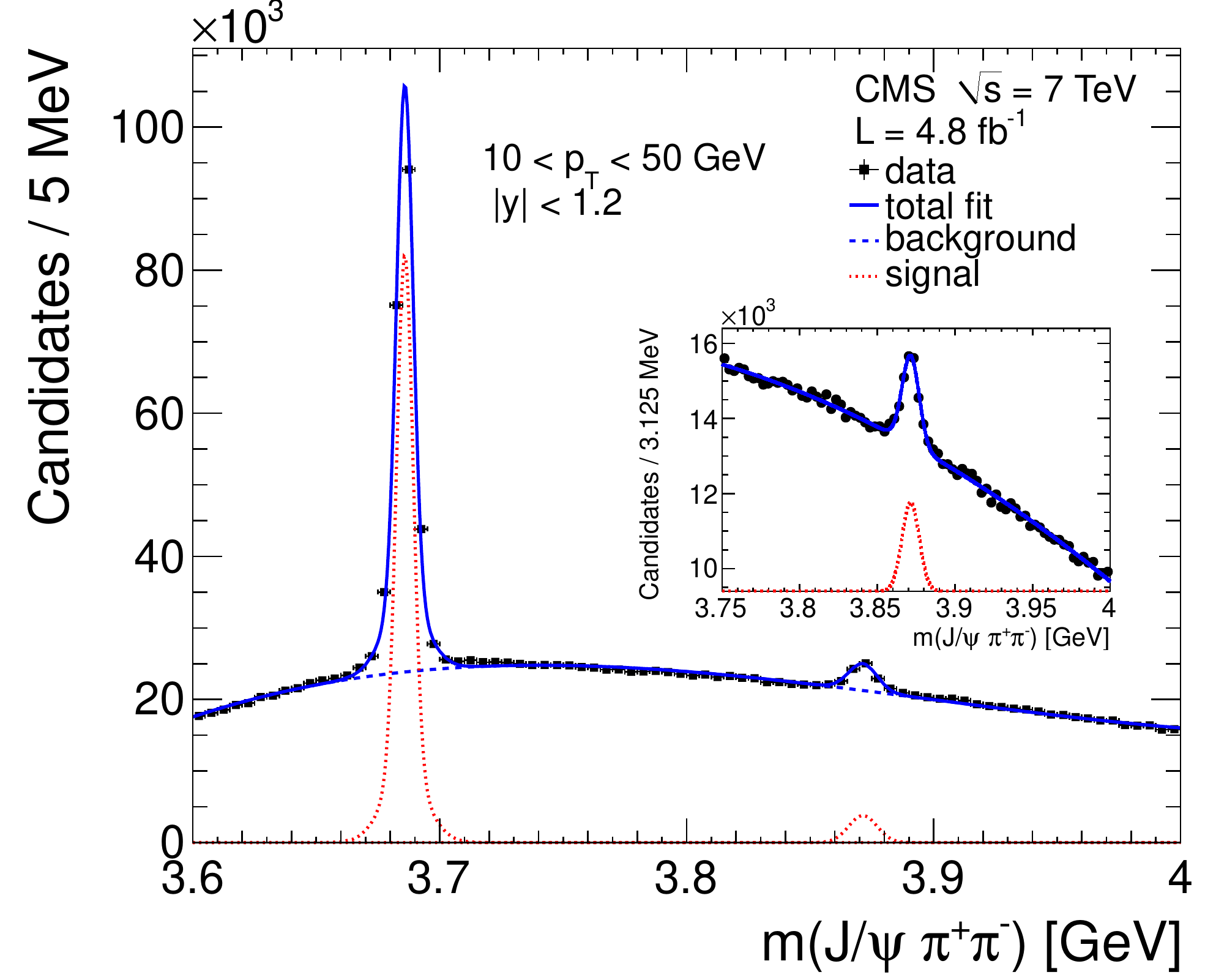}
\includegraphics[width=0.41\textwidth]{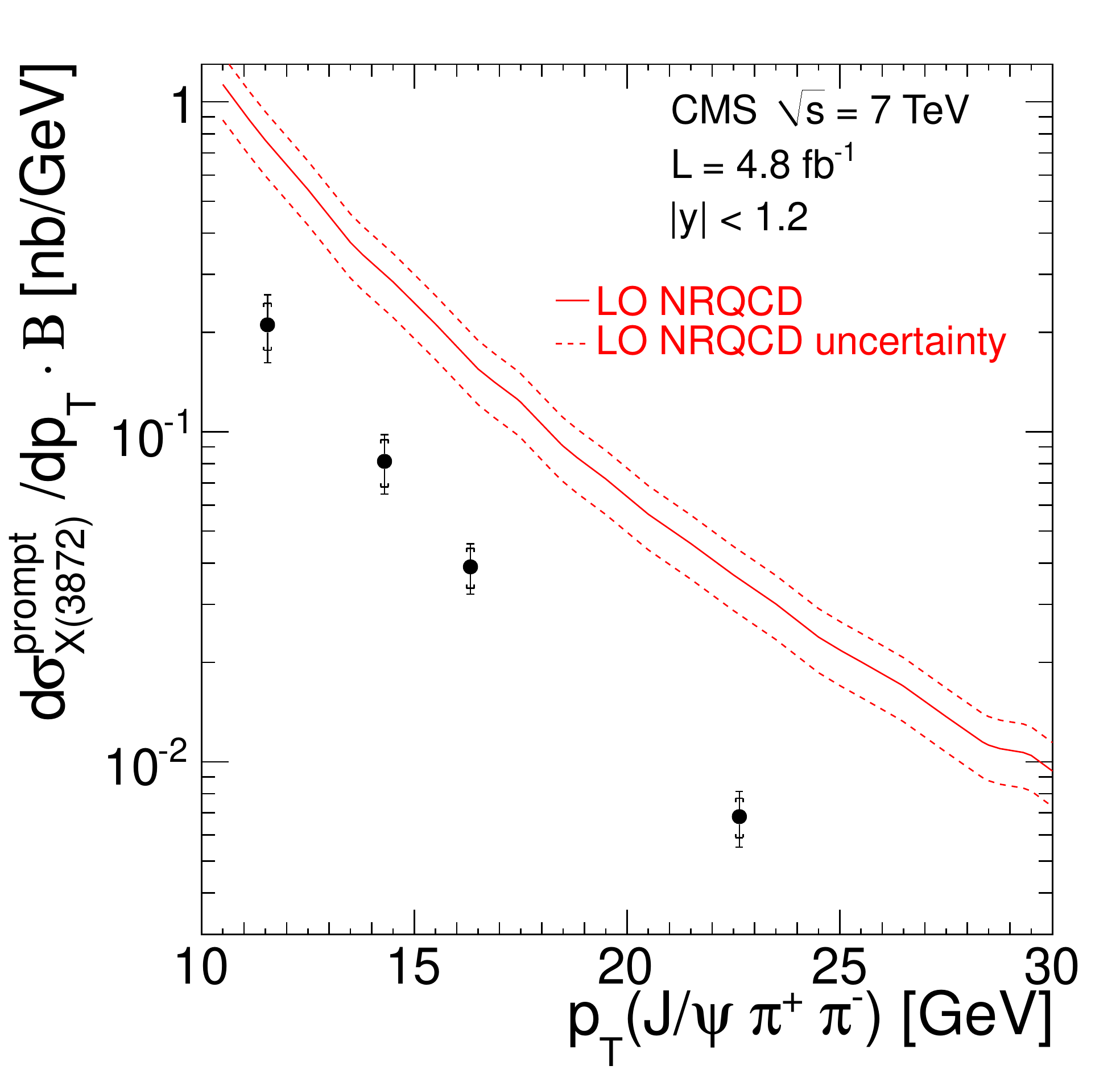}
\caption{\small 
  (Left) $\jpsi \pi^+ \pi^-$ mass spectrum observed by CMS (with $10 < \pt < 50 \gev$ and $|y| < 1.2$). 
  Clear signals for both the $X(3872)$ and $\psi(2S)$ states are seen.
  (Right) Differential cross-section for prompt $X(3872)$ versus $\pt$ compared to the NRQCD predictions~\cite{Artoisenet:2009wk}. 
  Both plots are taken from Ref.~\cite{Chatrchyan:2013cld}. 
}
\label{fig:xplots}
\end{figure}

The $X(3872)$ state has been studied in detail both at the $e^+e^-$ $B$ factories~\cite{Choi:2011fc,BaBarPhysRevD.71.071103} and the Tevatron~\cite{D0Abazov:2004kp,CDFPhysRevLett.93.072001} but the nature of this particle remains unclear. 
Its properties do not match the predictions for the conventional charmonium states and it has been interpreted as a candidate for a tetraquark state or a loosely bound deuteron-like $D^{*0}{\overline{D}}{}^0$ ``molecule'' (reviews can be found in Refs.~\cite{swanson,Godfrey:2008nc}). 
At the LHC the $X(3872)$ is produced both directly in $pp$ collisions and also in $\bquark$-hadron decays. 
Inclusive studies are challenging due to the large combinatorial background from other particles produced in the $pp$ interaction. 
Nevertheless, both LHCb and CMS have studied inclusive $X(3872)$ production in $pp$ collisions at $\sqrt{s} = 7 \tev$~\cite{LHCb-PAPER-2011-034,Chatrchyan:2013cld}. 
The cross-sections measured by both experiments are significantly less than those expected from leading-order NRQCD predictions~\cite{Artoisenet:2009wk}, as shown in Fig.~\ref{fig:xplots}.  

In addition, LHCb has measured the $X(3872)$ quantum numbers by performing a five-dimensional angular analysis of the $B^+ \rightarrow X(3872) K^+$, $X(3872) \to \jpsi \pi^+\pi^-$  decay chain~\cite{LHCb-PAPER-2013-001}. 
To distinguish between the possible hypotheses for the quantum numbers a likelihood ratio test is performed.  
The data favour the $J^{PC} = 1^{++}$ hypothesis and the $2^{-+}$ hypothesis is 
rejected with a significance of $8.4\sigma$. 
This assignment favours the hypothesis that the $X(3872)$ is exotic in nature.
However, the relative decay rates of the $X(3872)$ to the $\psi(2S)\gamma$ and $\jpsi\gamma$ final states~\cite{LHCb-PAPER-2014-008} are inconsistent with the prediction for a pure molecule.  
Further studies are still needed to understand the nature of the $X(3872)$ particle.

The large datasets collected by the LHC experiments have allowed studies
of other ``XYZ'' states. 
One very important result is the confirmation by LHCb~\cite{LHCb-PAPER-2014-014} of the existence and resonant nature of the $Z(4430)^+$ state first seen by the Belle collaboration~\cite{Choi:2007wga,Mizuk:2009da,Chilikin:2013tch}. 
As this state is charged its minimal quark content is $\cquark \cquarkbar \uquark \dquarkbar$, and thus it provides clear evidence for the existence of non-$q\overline{q}$ mesons.
A puzzle that remains open concerns the existence of the $X(4140)$ state. 
Evidence for this putative $\jpsi \phi$ resonance was reported by the CDF collaboration based on studies of the $B^+ \rightarrow \jpsi \phi K^+$ decay chain~\cite{Aaltonen:2009tz}. 
However, a subsequent LHCb study found no evidence for this state and set upper limits on its existence~\cite{LHCb-PAPER-2011-033}.  
An analysis by CMS~\cite{Chatrchyan:2013dma} confirmed the existence of a peaking structure in the same region, and in addition found evidence for a second structure with $m(\jpsi \phi) \sim 4300 \mevcc$, in same decay mode. 
Further study is needed to clarify whether these structures are resonant in nature.

\begin{figure}[!t]
\centering
\includegraphics[width=0.48\textwidth]{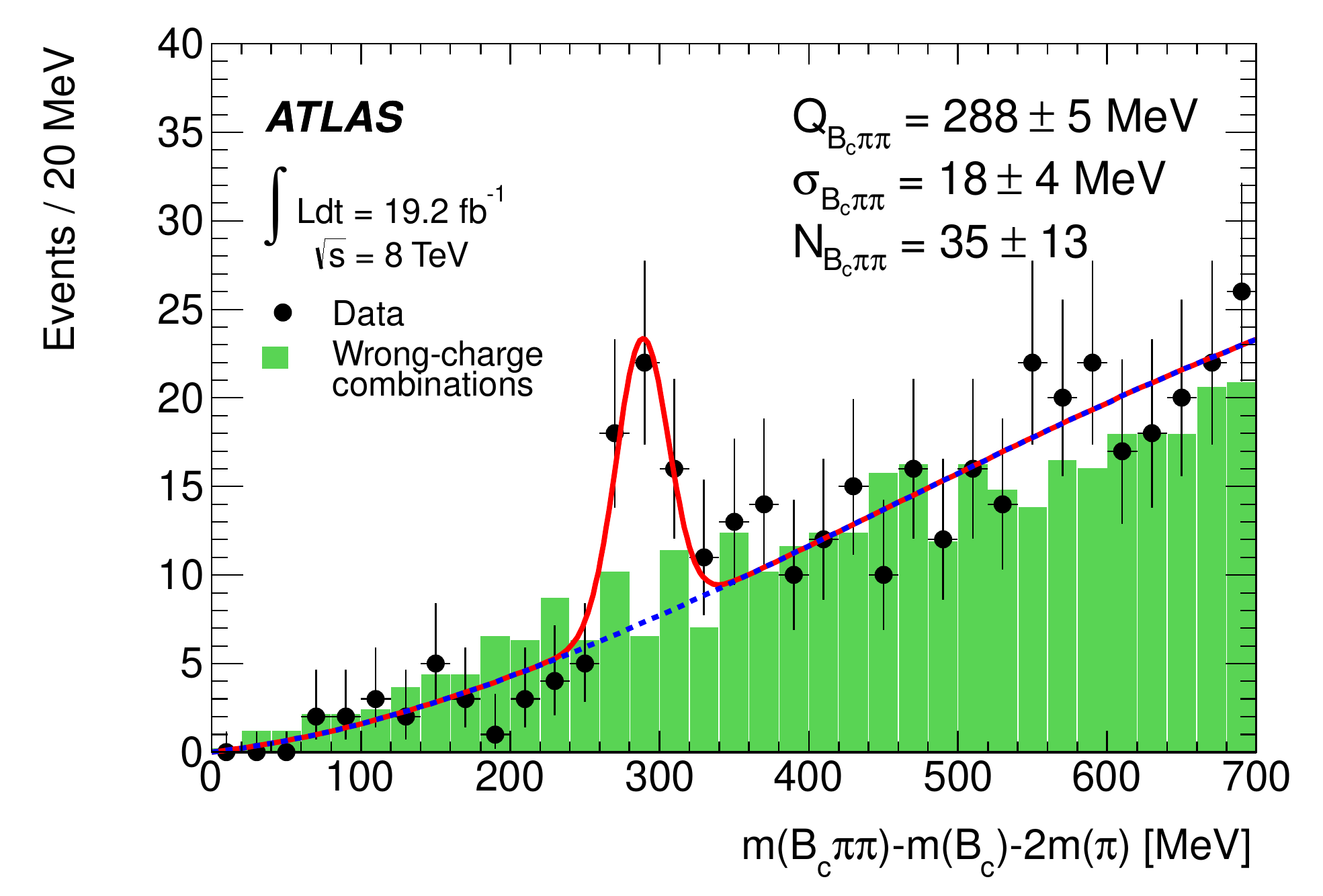}
\includegraphics[width=0.50\textwidth]{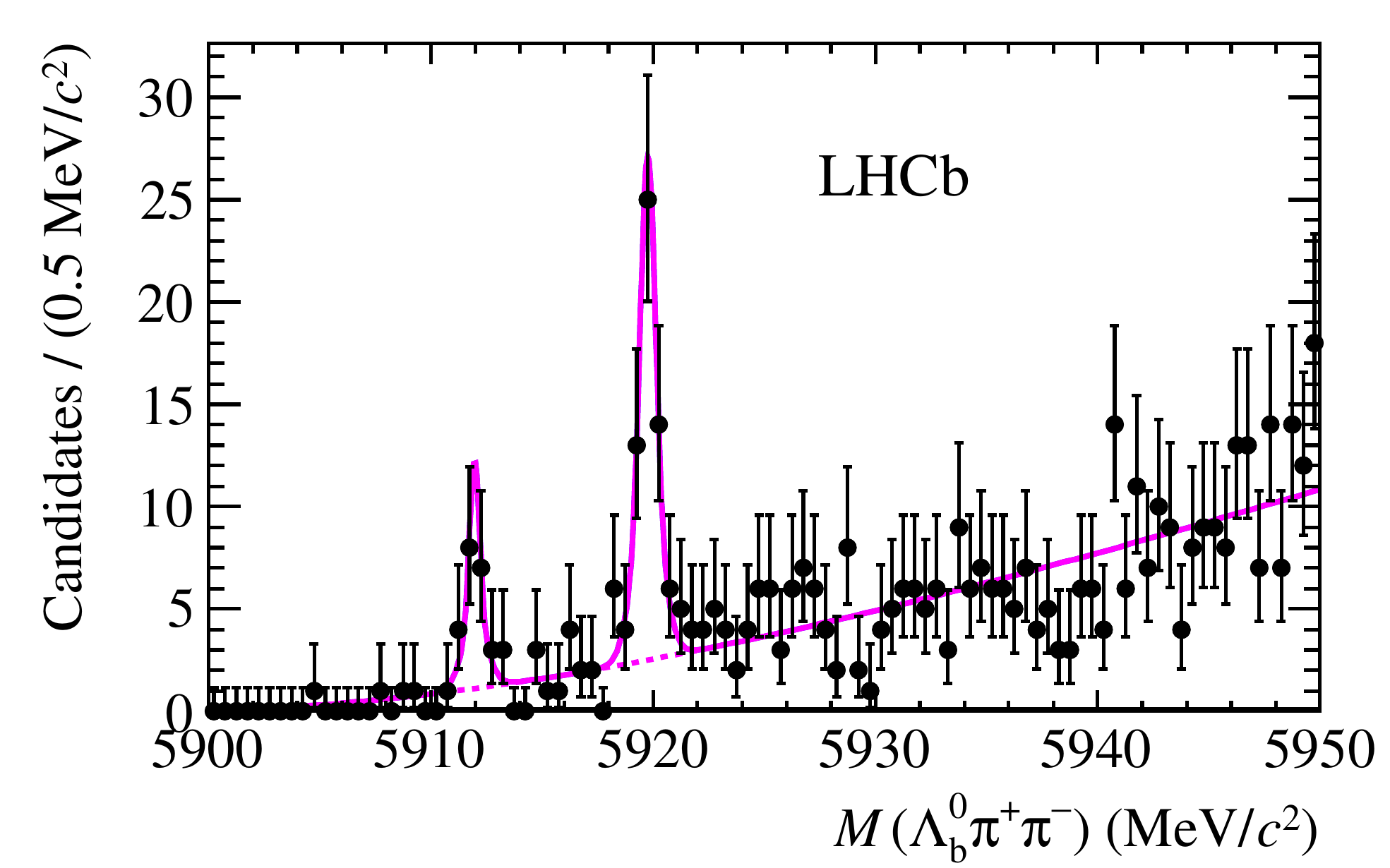}
\caption{\small
  Invariant mass distributions of: (left) $B_c^{\pm}\pip\pim$ candidates from ATLAS~\cite{Aad:2014laa}; (right) $\Lb \pip\pim$ candidates from LHCb~\cite{LHCb-PAPER-2012-012}.
}
\label{fig:bbaryons}
\end{figure}

The $B^{+}_{c}$ meson, the ground state of  the $\bquarkbar\cquark$~system, is unique as it is the only weakly decaying heavy quarkonium system. 
It was first observed by the CDF collaboration~\cite{Abe:1998wi,Abulencia:2005usa}, but with the advent of the LHC studies of the $B^+_c$ meson have entered a new precision era. 
LHCb has reported observations of many new decay modes~\cite{LHCb-PAPER-2011-044,LHCb-PAPER-2012-054,LHCb-PAPER-2013-010,LHCb-PAPER-2013-021,LHCb-PAPER-2013-047,LHCb-PAPER-2014-009},
while signals for the $B^{+}_{c} \rightarrow \jpsi \pi^{+} \pi^{+} \pi^{-}$ decay have also been reported by ATLAS~\cite{ATLAS-CONF-2012-028} and CMS~\cite{Khachatryan:2014nfa}. 
One particularly important observation is that of the decay $B_c^+ \rightarrow B_s^0\pi^+$~\cite{LHCb-PAPER-2013-044}, which is mediated by the decay of the constituent $\cquark$ quark.
Such decays are expected to dominate the $B_c^+$ width, but had never previously been observed.
It will be of interest to search for decay modes in which the $\bquarkbar$ and $\cquark$ annihilate in the near future.
The fundamental properties of the $B_c^+$ meson have also been determined, with significant improvements in precision compared to prior results. 
The most precise measurement of the $B_c^+$ mass to date is obtained by LHCb in the $\jpsi D_s^+$ decay mode~\cite{LHCb-PAPER-2013-010}, 
$$m(B^+_c) = 6276.28 \pm 1.44 \stat \pm 0.36 \syst \mevcc \, .$$
In addition, LHCb has used semileptonic $B_c^+$ decays to measure the lifetime~\cite{LHCb-PAPER-2013-063}
$$\tau(B_c^+) = 509 \pm 12 \stat \pm 8 \syst \fs \, .$$
The observation of a candidate $B_c(2S)^+$ state by ATLAS~\cite{Aad:2014laa}, shown in Fig.~\ref{fig:bbaryons}(left), demonstrates the possibility for more detailed understanding of the spectroscopy in the \Bc sector.


The large data samples collected by the LHC experiments has also allowed to explore in detail the properties of the $b$-baryon sector for the first time. 
One puzzle from studies at LEP and the Tevatron concerned the lifetime of the $\Lb$ baryon. 
Theoretical predictions based on the Heavy Quark Expansion (HQE) give the ratio of the $\Lb$ and $\Bz$ lifetimes to be consistent with unity at the level of a few percent~\cite{PhysRevD.56.2783,Neubert:1996we,Uraltsev:1996ta}. 
However, early measurements gave considerably smaller values.
ATLAS, CMS and LHCb have all made measurements of this quantity~\cite{Aad:2012bpa,Chatrchyan:2013sxa,LHCb-PAPER-2013-032} and find values more consistent with theory. 
The most precise measurements of the $\Lb$ lifetime are obtained by LHCb with decays to the $\jpsi \Lz$~\cite{LHCb-PAPER-2013-065} and $\jpsi \proton K^-$~\cite{LHCb-PAPER-2014-003} final states.
These give
$$\tau(\Lb) = 1.468 \pm  0.009 \pm 0.008 \ps\,,$$ 
within a few percent of the measured $\Bz$ lifetime~\cite{PDG2012} as predicted by the HQE. 
Measurements of the lifetimes of other $\bquark$-baryons also start to approach the percent level~\cite{LHCb-PAPER-2014-010,LHCb-PAPER-2014-021}.

First observations of excited $b$-baryons have also been made. 
CMS has observed an excited $\Xires_b$ state~\cite{Chatrchyan:2012ni} whilst LHCb has observed two excited $\Lb$ states~\cite{LHCb-PAPER-2012-012}, as shown in Fig.~\ref{fig:bbaryons}(right).
Further new discoveries in this area can be anticipated with more data and refined analysis techniques. 

\section{Mixing}
\label{sec:mixing}

There are four ground-state neutral flavoured mesons: the \Kz ($\bar{s}d$), \Dz ($c\bar{u}$), \Bz ($\bar{b}d$) and \Bs  ($\bar{b}s$) particles.
Each of these can mix with its antiparticle, through diagrams that either contain virtual heavy particles or have on-shell intermediate states.
These are often referred to as short-distance (or dispersive) and long-distance (or absorptive) processes, respectively.

As a result of the mixing, physical states with definite masses and lifetimes are formed.
When the amplitudes of the mixing processes are calculated, theoretical predictions for the mass and width differences ($\dm$ and $\DG$), and for the parameter of \CP violation in the mixing ($a_{\rm sl}$), are obtained.
Since, at least for the short-distance diagrams, the uncertainties related to the theory prediction are under good control, comparison of the measurements (particularly $\Delta m$ and $a_{\rm sl}$) to the predictions provide strong tests of the Standard Model.

The first data from the LHC has led to significant improvement in knowledge of the \Dz and \Bs mixing parameters.
The measurement of the mass difference in the \Bs system, $\dms$, had proved challenging for previous experiments because its large value causes fast oscillations between \Bs and \Bsb that are hard to resolve.
The CDF collaboration did, however, succeed to determine the value of $\dms$ with $5\sigma$ significance~\cite{Abulencia:2006ze}.

\begin{figure}[!t]
\centering
\includegraphics[width=0.58\textwidth]{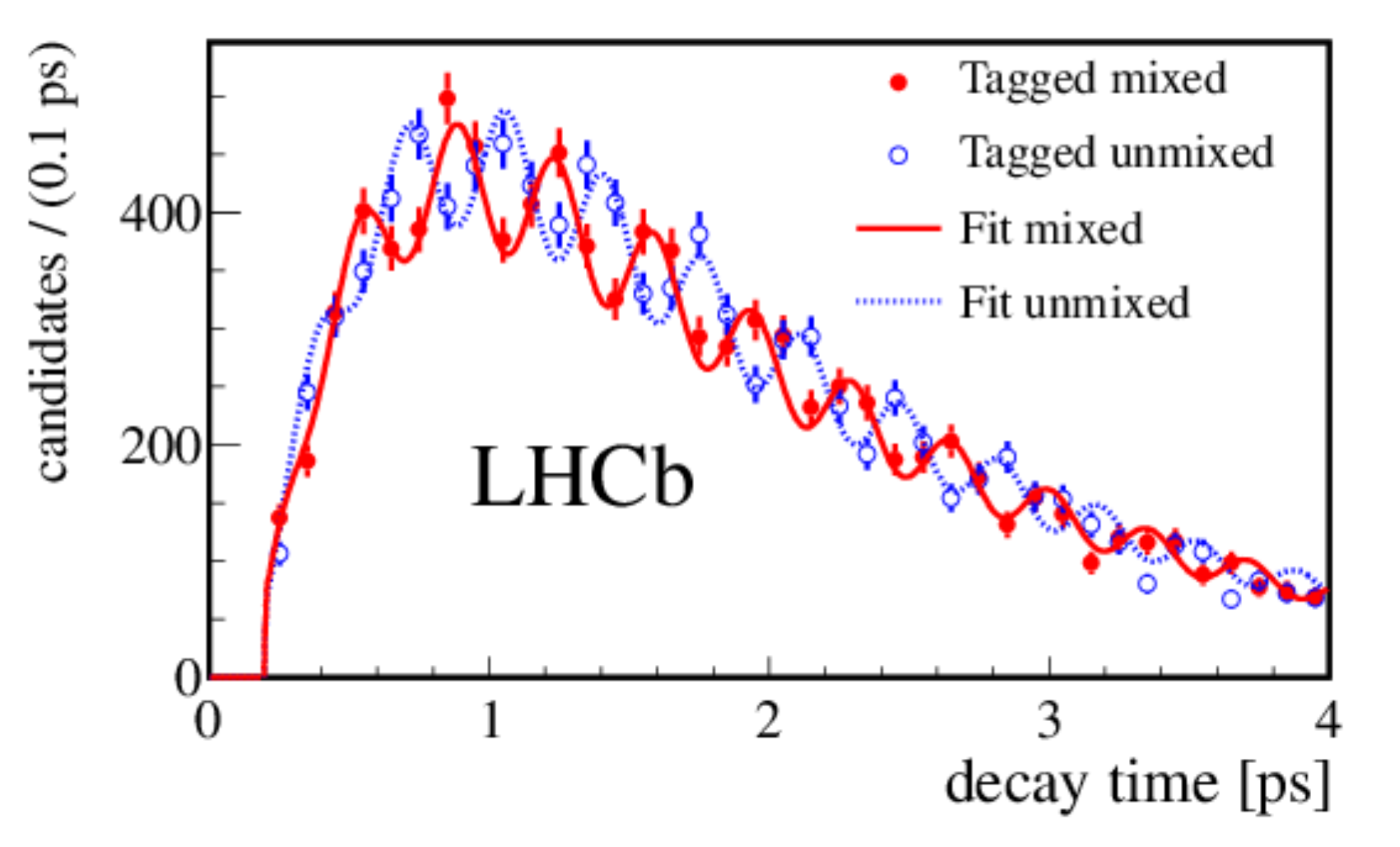}
\includegraphics[width=0.38\textwidth]{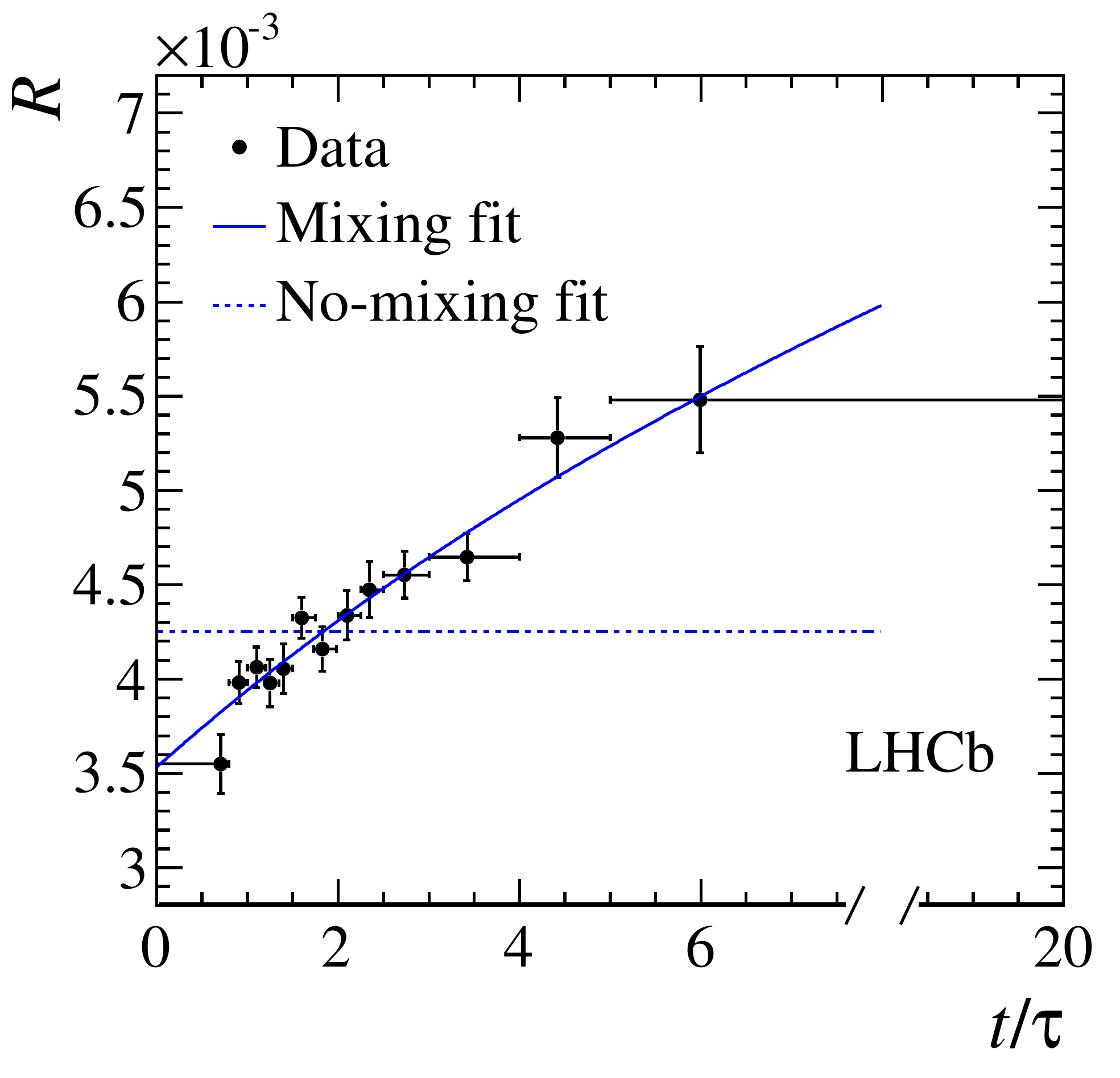}
\caption{\small
  Mixing in (left) the $\Bs$--$\Bsb$ system~\cite{LHCb-PAPER-2013-006}; (right) the $\Dz$--$\Dzb$ system~\cite{LHCb-PAPER-2012-038}.
}
\label{fig:mix}
\end{figure}

Due to the LHCb detector's excellent vertex resolution, which allows to resolve the oscillations, and flavour tagging capability~\cite{LHCb-PAPER-2011-027,LHCb-CONF-2012-033}, from which initial state \Bs and \Bsb mesons can be distinguished, major improvement in the determination of \dms has been achieved~\cite{LHCb-PAPER-2011-010,LHCb-PAPER-2013-006,LHCb-PAPER-2013-036}.
The single most precise result, based on a sample of $\Bs \to \Dsm\pip$ decays selected from LHCb's 2011 data sample, illustrated in Fig.~\ref{fig:mix}~(left), gives~\cite{LHCb-PAPER-2013-006}
$$
\dms = ( 17.768 \pm 0.023 \stat \pm 0.006 \syst ) \invps \, .
$$

The width difference in the \Bs system, \DGs, can be measured either by comparing the lifetimes for \CP-even (\eg\ $\Kp\Km$~\cite{LHCb-PAPER-2011-014,LHCb-PAPER-2012-013,LHCb-PAPER-2014-011} or $\Dsp\Dsm$~\cite{LHCb-PAPER-2013-060}) and \CP-odd (\eg\ $\jpsi \pip\pim$~\cite{LHCb-PAPER-2012-017} or $\jpsi \KS$~\cite{LHCb-PAPER-2013-015}) final states or from analysis of decays to a final state that contains an admixture of both (\eg\ $\jpsi\phi$).
The most precise measurements come from the latter approach~\cite{Aad:2012kba,Aad:2014cqa,CMS-BPH-13-012,LHCb-PAPER-2011-021,LHCb-PAPER-2013-002}, which also has the advantage that it is not necessary to make assumptions concerning \CP violation parameters since they can be determined simultaneously (as discussed below).
The single most precise measurement gives~\cite{LHCb-PAPER-2013-002}
$$
\DGs = ( 0.100 \pm 0.016 \stat \pm 0.003 \syst ) \invps \, .
$$
In this channel, by studying the variation of the strong phase difference between the $\jpsi\phi$ and $\jpsi \Kp\Km$ (S wave) components, the sign of \DGs has been confirmed to be positive~\cite{LHCb-PAPER-2011-028,LHCb-PAPER-2013-002}.

In stark contrast to the large values of \dms and \DGs, the mixing parameters in the charm sector are small.
Indeed, while previous experiments had seen evidence for charm oscillations~\cite{Aubert:2007wf,Staric:2007dt,Aaltonen:2007ac}, no single measurement exceeded the $5\sigma$ threshold for discovery.
LHCb has taken the precision of the measurements far past that threshold using $\Dz\to\Kpm\pimp$ decays~\cite{LHCb-PAPER-2012-038,LHCb-PAPER-2013-053} as shown in Fig.~\ref{fig:mix}~(right).
The world averages of the charm mixing parameters are now~\cite{hfag}
$$
x_D = \dm_D/\Gamma_D = (0.39\,^{+0.16}_{-0.17})\% \, , ~~~~~
y_D = \DG_D/(2\Gamma_D) = (0.67\,^{+0.07}_{-0.08})\% \, ,
$$
where $\Gamma_D$ is the average width of the neutral charm mesons.
As the value of $x_D$ is still consistent with zero, further improvement of precision is well motivated.

\section{Mixing-related \CP violation}
\label{sec:mixCPV}

\CP violation phenomena in charm oscillations are expected to be negligible.
Now that charm mixing is definitively established, precise experimental searches to test this Standard Model prediction are imperative.  
Results from LHCb with the $\Dz \to \Kmp\pipm$~\cite{LHCb-PAPER-2013-053} and $\Dz\to\Kp\Km$ and $\Dz\to\pip\pim$~\cite{LHCb-PAPER-2013-054} decays have significantly improved the precision of prior measurements, but not yet revealed any discrepancy with the Standard Model.
Further improvements in precision are anticipated as all measurements are updated to the full Run~I data sample, and results from additional channels such as $\Dz \to \KS\pip\pim$ become available.

Mixing-induced \CP violation in the $\Bz$ sector has been studied extensively by the $\epem$ $B$ factories.
The benchmark measurement is from the asymmetry in $\Bz \to \jpsi \KS$ decays, which gives $\sin2\beta = 0.665 \pm 0.024$~\cite{Aubert:2009aw,Adachi:2012et,hfag}, where $\beta \equiv {\rm arg}\left[ -V^{}_{cd}V_{cb}^* /(V^{}_{td}V_{tb}^*) \right]$ and $V^{}_{ij}$ represent elements of the CKM matrix.
Using its 2011 dataset LHCb has measured $\sin2\beta = 0.73 \pm 0.07 \stat \pm 0.04 \syst$~\cite{LHCb-PAPER-2012-035}, in agreement with the world average though less precise. 
Results with the full Run~I precision should be competitive with the individual measurements from BaBar and Belle.

The LHC era has allowed the first high precision searches for \CP violation in the $\Bs$ sector. 
In the Standard Model the prediction for the \CP violation effect in $\Bs\to\jpsi\phi$ decays is $\phi_s \equiv -2 {\rm arg} \left[ -V^{}_{ts}V_{tb}^* /(V^{}_{cs}V_{cb}^*) \right] = -0.036 \pm 0.002$. 
Extensions of the Standard Model typically give additional phases which can make the measured value of $\phi_s$ differ from this prediction. 
Hence, precise measurements of $\phi_s$ test models of new physics. 
Prior to LHC running, first measurements from the Tevatron experiments hinted towards a deviation with the Standard Model, though with large uncertainties~\cite{Aaltonen:2007he,Abazov:2008af}.
Significant improvement in precision has been achieved by LHCb with the $\Bs \to \jpsi\phi$ channel~\cite{LHCb-PAPER-2011-021,LHCb-PAPER-2013-002}.
ATLAS~\cite{Aad:2014cqa} and CMS~\cite{CMS-BPH-13-012} have also released preliminary measurements from flavour-tagged time-dependent angular analyses of $\Bs \to \jpsi\phi$ decays, using their 2011 and 2012 data samples, respectively.

\begin{figure}[!t]
\centering
\includegraphics[width=0.7\textwidth,bb=100 100 680 495,clip=true]{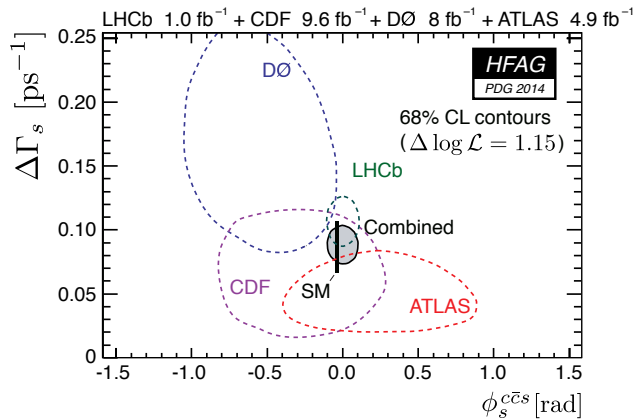}
\caption{\small 
  Compilation of data on $\phi_s$ and $\DGs$~\cite{hfag}.
  The latest results from LHCb~\cite{LHCb-PAPER-2014-019} and CMS~\cite{CMS-BPH-13-012} are not included.
}
\label{fig:phis}
\end{figure}

It is of interest to make $\phi_s$ measurements using additional channels.
LHCb has released results based on the $\Bs \to \jpsi \pip\pim$ decay~\cite{LHCb-PAPER-2011-031,LHCb-PAPER-2012-006,LHCb-PAPER-2014-019}.
This final state has a large contribution from $\jpsi f_0(980)$ decays, but the whole phase-space can be used inclusively since it has been shown to be dominantly \CP-odd~\cite{LHCb-PAPER-2012-005,LHCb-PAPER-2013-069}.
The most recent LHCb analysis of $\Bs \to \jpsi \pip\pim$ using the full Run~I data sample gives currently the single most precise measurement~\cite{LHCb-PAPER-2014-019}
$$\phi_s =  0.07 \pm 0.07 \stat \pm 0.01 \syst \ {\rm rad}  \, .$$
Further improvement in precision is anticipated with the update of the $\Bs \to \jpsi \phi$ analysis to the full Run~I sample.
Figure~\ref{fig:phis} summarises the current knowledge of $\phi_s$: the measurements are consistent with the Standard Model, suggesting that any new physics effects are small. 

In addition, LHCb has measured $\phi_s$ with the $\Bs \to \phi\phi$ channel~\cite{LHCb-PAPER-2013-007,LHCb-PAPER-2014-026}.
As this decay is dominated by the $b \to s$ loop transition, it can be used to search for new physics effects in the decay amplitude as well as in mixing.
In the near future it will be interesting to obtain also values of $\phi_s$ from $\Bs \to \Kstarz \Kstarzb$ which is similarly sensitive to possible new physics effects.

\begin{figure}[!t]
\centering
\includegraphics[width=0.5\textwidth]{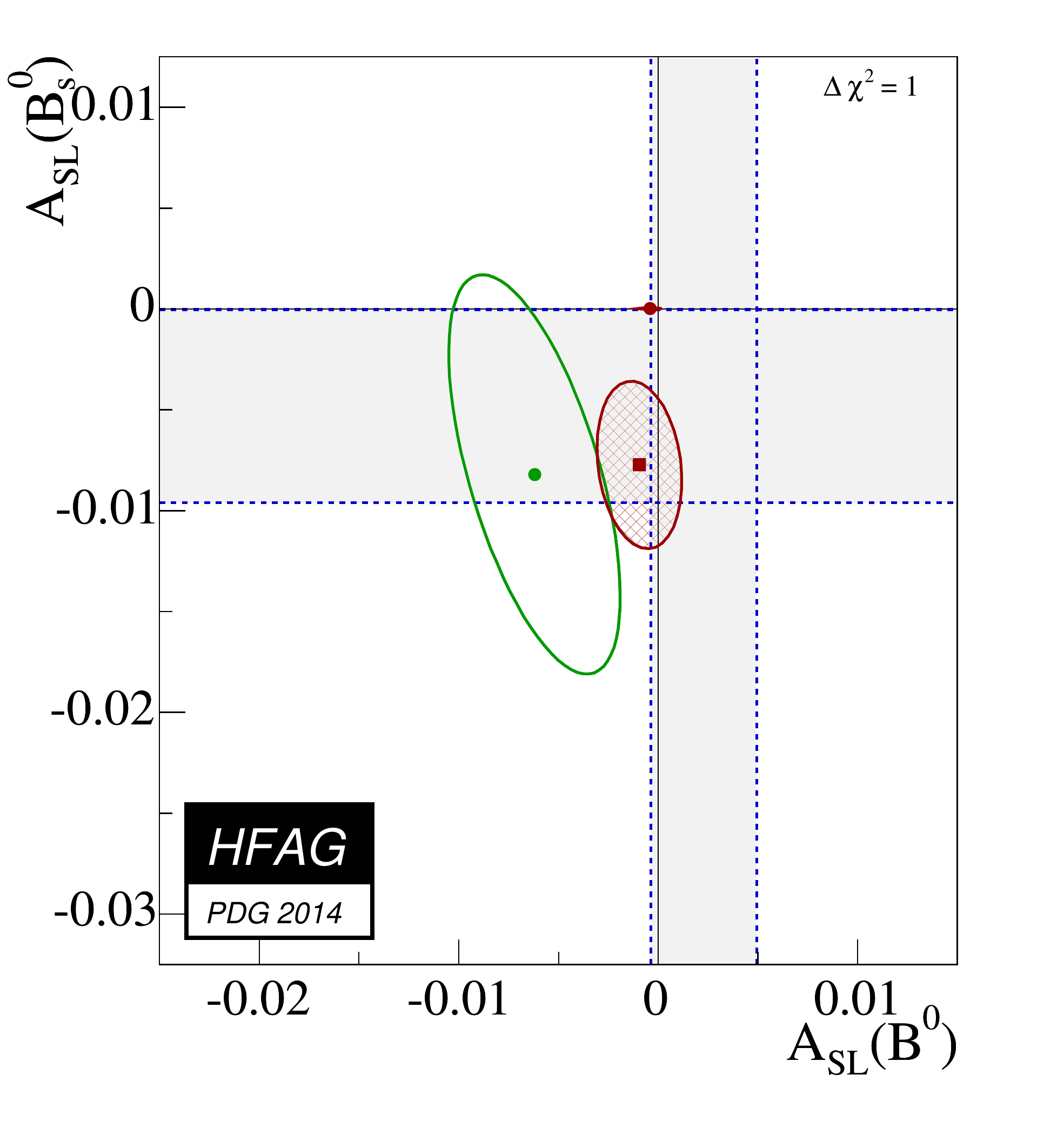}
\caption{\small 
  Compilation of data on $a_{\rm sl}$ in the $\Bz$ and $\Bs$ system~\cite{hfag}. 
  The vertical and horizontal bands show the averages of $a_{\rm sl}(\Bz)$ and $a_{\rm sl}(\Bz)$ measured individually, the green ellipse is the D0 measurement with inclusive same-sign dileptons, and the red ellipse is the world average of all measurements.
  The Standard Model prediction (red point) is indistinguishable from the origin.
}
\label{fig:asl}
\end{figure}

The semileptonic (or flavour-specific) asymmetry $a_{\rm sl}$ is sensitive to \CP violation in mixing.
It is expected to be negligibly small in the Standard Model, but the D0 collaboration has reported an anomalous asymmetry between positive and negative like-sign muon pairs~\cite{Abazov:2013uma}. 
The effect, which deviates from the Standard Model predictions by $3.6\sigma$, could be caused by non-zero $a_{\rm sl}$ in either or both the \Bd and \Bs systems.
LHCb has measured $a_{\rm sl}(\Bs)$ using $\Bs \to \Dsm\mup X$ decays reconstructed in its 2011 data sample~\cite{LHCb-PAPER-2013-033}.
The result is the single most precise measurement of this quantity, and is in agreement with the Standard Model prediction, as are all other measurements of $a_{\rm sl}(\Bd)$ and $a_{\rm sl}(\Bs)$ individually. 
The current situation is summarised in Fig.~\ref{fig:asl}. 
Further improvements in precision are needed to shed light on whether the D0 result indicates new physics.

\section{Direct \CP violation and the determination of $\gamma$}
\label{sec:directCPV}

The term ``direct \CP violation'' is used to refer to asymmetries that cannot be caused by \CP violation in the mixing amplitude.
Historically, this categorisation was important to test so-called ``superweak'' models.
Direct \CP violation can be observed as a difference in mixing-related \CP violating effects in neutral meson decays to different final states, as seen in the kaon system~\cite{Fanti:1999nm,AlaviHarati:1999xp}.
Alternatively, if $a_{\rm sl}$ is known to be small, direct \CP violation can be seen in flavour-specific decays of neutral mesons, for example as an asymmetry in the yields of $\Bd \to \Kp\pim$ and $\Bzb \to \Km\pip$ decays~\cite{Aubert:2004qm,Chao:2004mn}.
This approach allowed LHCb to make the first observation of \CP violation in the \Bs system through the decay rate asymmetry of $\Bs\to\Km\pip$ and $\Bsb\to\Kp\pim$ decays,~\cite{LHCb-PAPER-2011-029,LHCb-PAPER-2013-018}
$$
A_{\CP}(\Bs\to\Km\pip) = 0.27 \pm 0.04 \stat \pm 0.01 \syst \, .
$$

\begin{figure}[!t]
\centering
\includegraphics[width=0.48\textwidth]{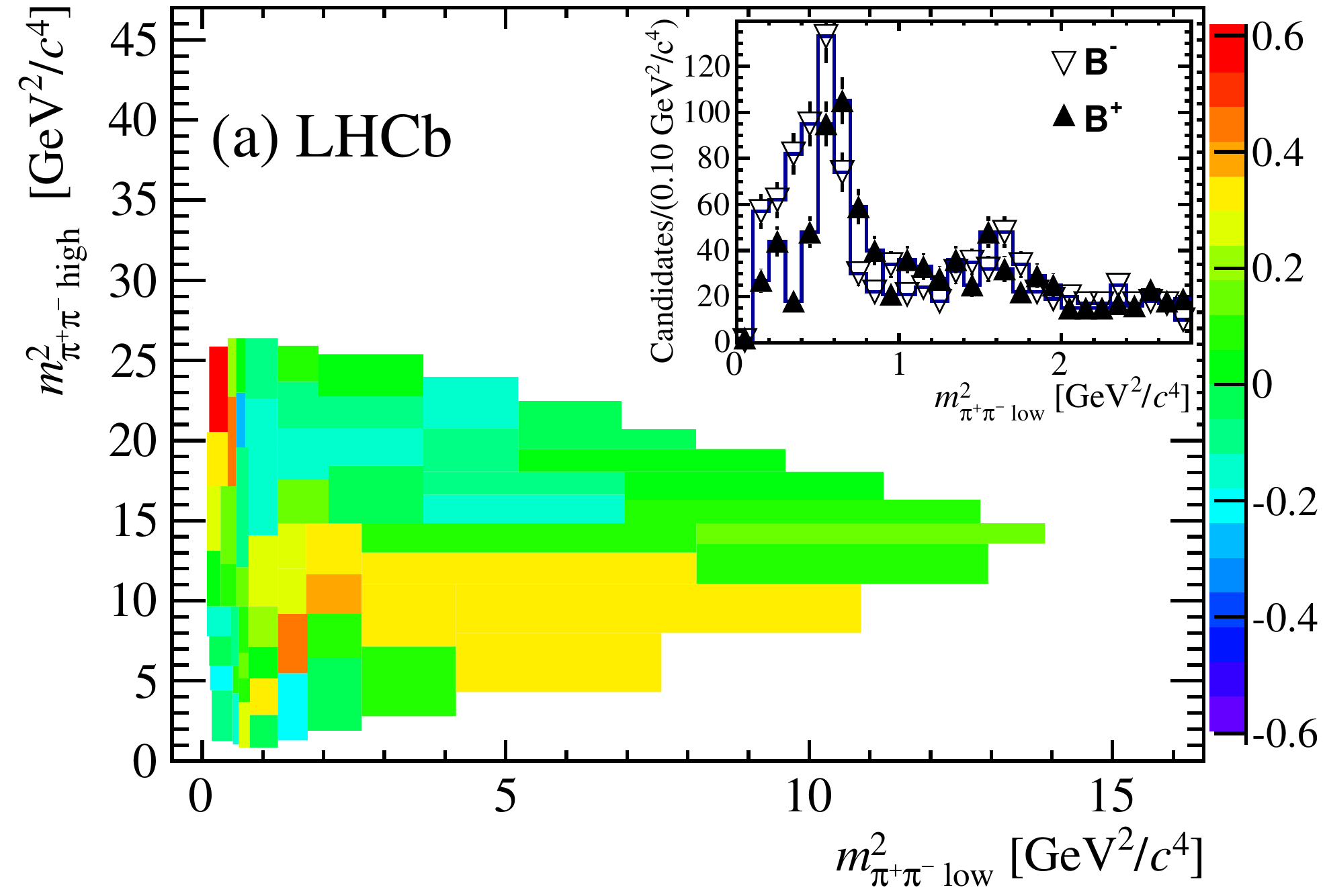}
\includegraphics[width=0.48\textwidth]{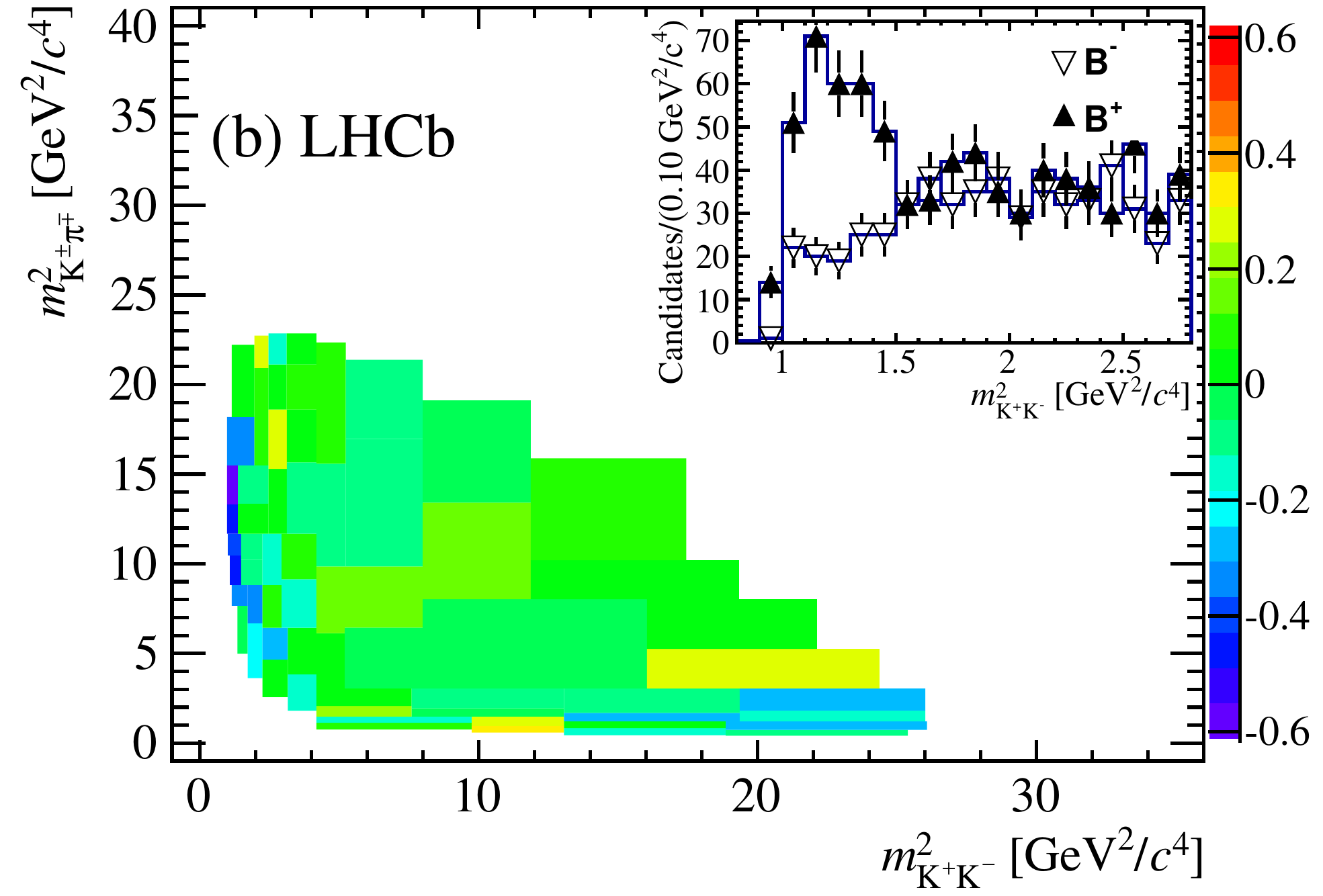}
\caption{\small
  Direct \CP violation effects in the phase space of (left) $\Bp\to\pip\pim\pip$ and (right) $\Bp\to\Kp\Km\pip$ decays~\cite{LHCb-PAPER-2013-051}.
  The $z$-axis ($y$-axis on inset) shows the raw asymmetry, without correction for background, or instrumental efficiencies and asymmetries.
}
\label{fig:dcpv}
\end{figure}

Among other results on direct \CP violation, particularly striking are the phase-space dependent effects seen in $\Bp$ decays to the final states $\pip\pim\Kp$, $\Kp\Km\Kp$, $\pip\pim\pip$ and $\Kp\Km\pip$~\cite{LHCb-PAPER-2013-027,LHCb-PAPER-2013-051}, illustrated in Fig.~\ref{fig:dcpv}.
The asymmetries are larger than 50\,\% in some regions away from resonant peaks.
This appears to indicate that interference effects play a crucial role in generating the asymmetries, although further investigation is necessary to obtain a clear understanding.

The main objective in \CP violation studies is to understand whether all observed effects arise from the complex phase of the CKM quark mixing matrix.
To interpret results such as those in charmless $B$ meson decays mentioned above, it is necessary to have a benchmark determination of the Standard Model phase $\gamma \equiv {\rm arg}\left[ -V^{}_{ud}V_{ub}^*/(V^{}_{cd}V_{cb}^*) \right]$.
This can be achieved using decays such as $B \to DK$, where interference between amplitudes leading to final state \Dz and \Dzb mesons is possible when they are reconstructed in a common final state such as $\Kp\Km$.  
As no loop contributions are possible these decays are very clean theoretically, and moreover all hadronic parameters can be determined from the data by considering several different $D$ meson decays.
LHCb has reported results on $\Bp \to \D \Kp$ using $\D \to \Kp\Km, \pip\pim, \Kmp\pipm$~\cite{LHCb-PAPER-2012-001}, $\Kmp\pipm\pip\pim$~\cite{LHCb-PAPER-2012-055}, $\KS\pip\pim, \KS\Kp\Km$~\cite{LHCb-PAPER-2012-027,LHCb-PAPER-2014-041} and $\KS\Kpm\pimp$~\cite{LHCb-PAPER-2013-068} decays.
The combination of these results gives~\cite{LHCb-PAPER-2013-020,LHCb-CONF-2013-006} 
$$
\gamma = (67 \pm 12)^\circ \, .
$$
Since most of the inputs are yet to be updated to include all Run~I data, and results on $\Bz \to \D \Kstarz$~\cite{LHCb-PAPER-2012-042,LHCb-PAPER-2014-028}, $\Bp \to \D \Kp \pip\pim$~\cite{LHCb-CONF-2012-021} and $\Bs \to \Dsmp\Kpm$~\cite{LHCb-PAPER-2014-038} can also be included, there are excellent prospects for further improvement.

Since the Standard Model predicts only small or vanishing \CP asymmetries in charm decays, searches for direct \CP violation provide useful null tests.
A measurement of the parameter $\Delta A_{\CP}$, describing the difference between the asymmetries for $\Dz \to \Kp\Km$ and $\Dz \to \pip\pim$ decays, that differed from zero by 3.5 standard deviations~\cite{LHCb-PAPER-2011-023}, prompted a great deal of theoretical activity (reviewed in Ref.~\cite{LHCb-PAPER-2012-031}) to examine whether or not a percent level asymmetry could be accommodated within the Standard Model.
More recent measurements, however, bring the world average closer to zero~\cite{LHCb-CONF-2013-003,LHCb-PAPER-2013-003,LHCb-PAPER-2014-013}. 
In addition searches for \CP violation in \Dp and \Dsp decays have yielded null results~\cite{LHCb-PAPER-2011-017,LHCb-PAPER-2012-052,LHCb-PAPER-2013-041,LHCb-PAPER-2013-057,LHCb-PAPER-2014-018}, so that currently there is no evidence for \CP violation in the charm system.

\section{Rare decays}
\label{sec:rare}

A powerful way to search for new physics is to study decays which are either forbidden or suppressed due to features of the Standard Model that may not be present in a more general theory.
The flavour sector of the Standard Model presents several such features: there are no flavour changing neutral currents at tree-level, vertices involving quarks of different families are suppressed by CKM quark mixing matrix elements, and the $V-A$ structure of the weak interaction leads to distinctive effects.
Rare decays of \B mesons to final states containing leptons or photons are particularly useful for Standard Model tests since observables can be calculated with reduced theoretical uncertainty compared to fully hadronic final states. 

\begin{figure}[!t]
\centering
\includegraphics[width=0.54\textwidth]{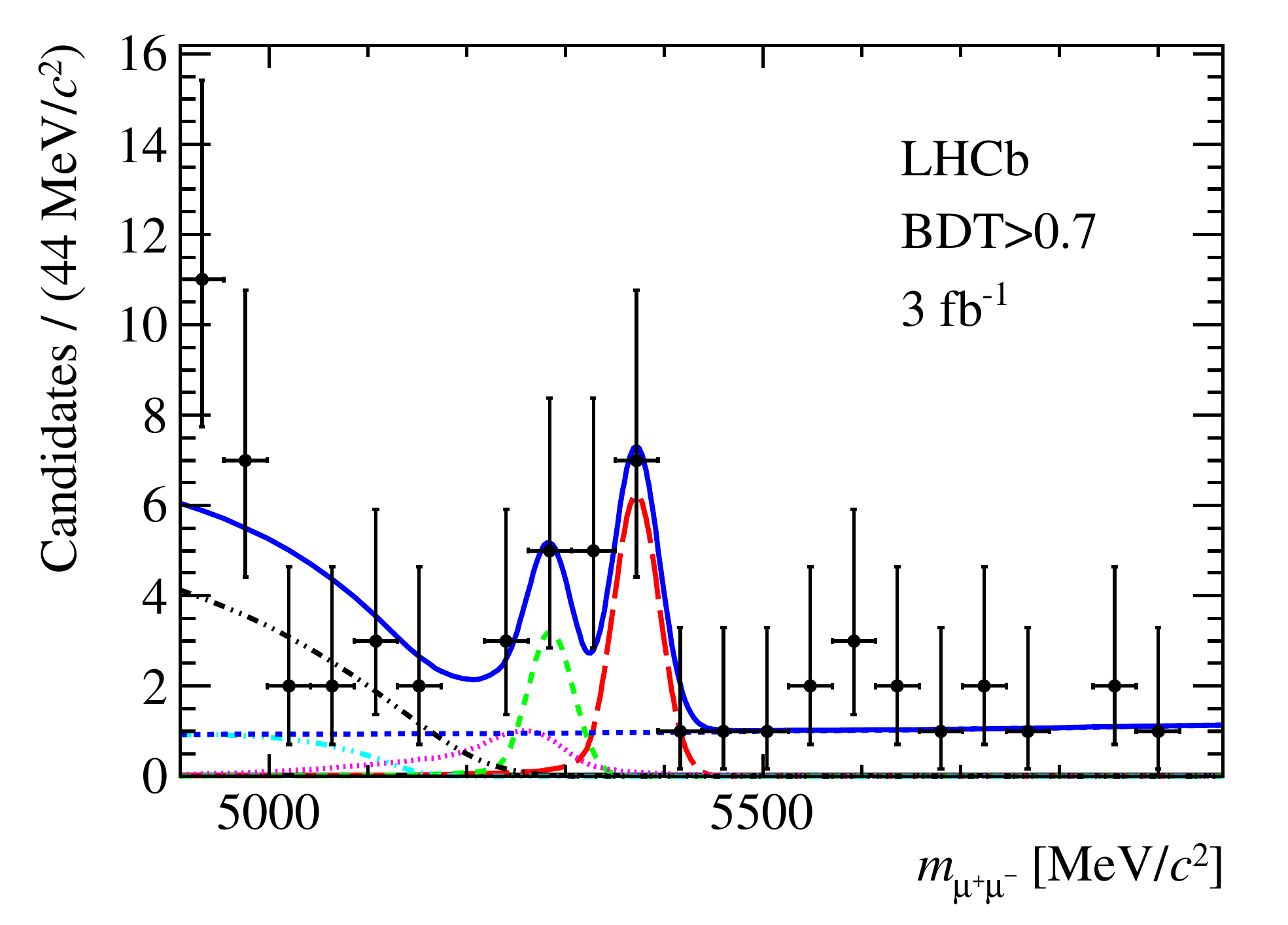}
\includegraphics[width=0.45\textwidth]{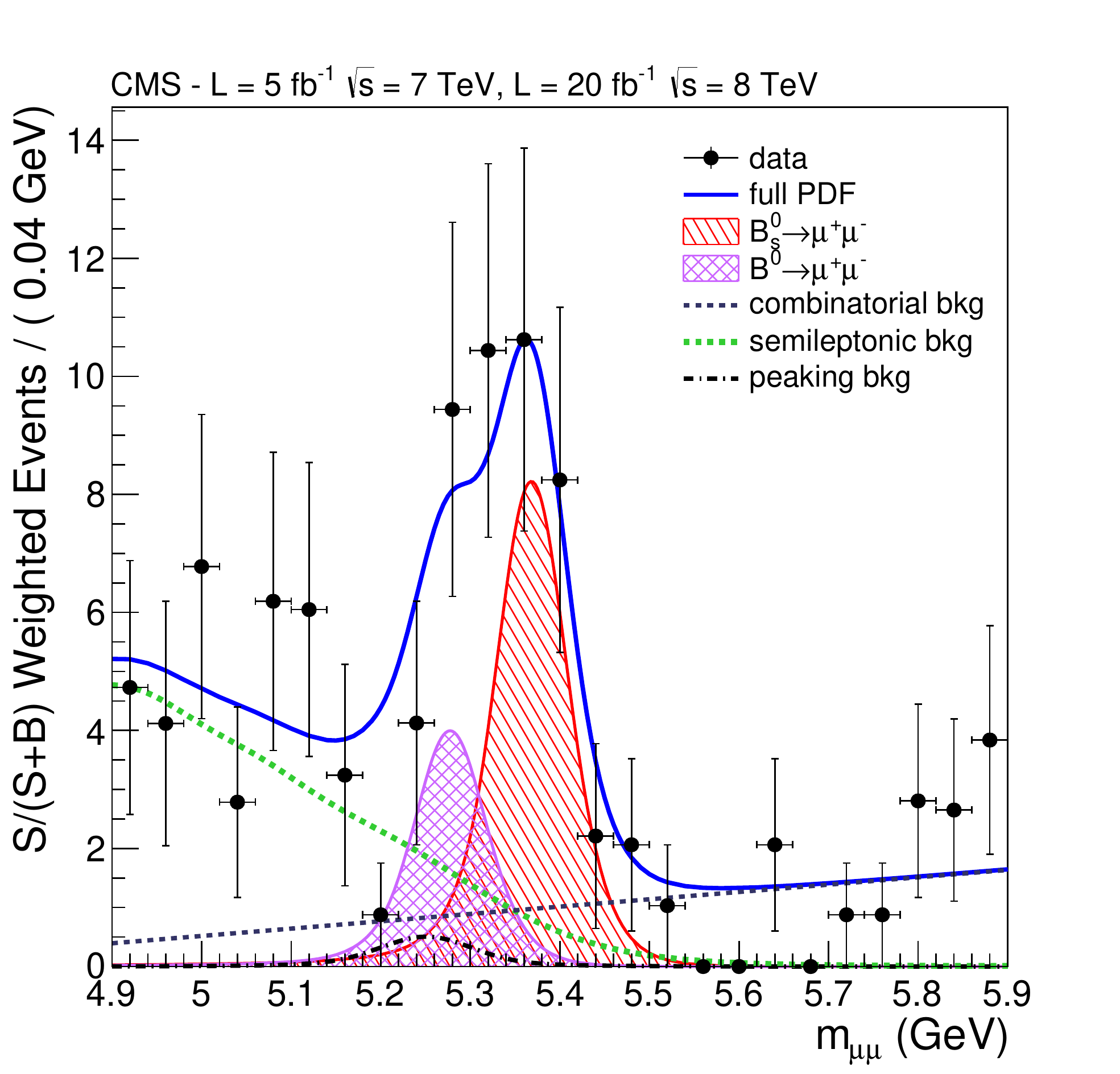}
\caption{\small
  Signals for $\Bs \to \mumu$ decays from (left) LHCb~\cite{LHCb-PAPER-2013-046} with only the most signal-like candidates included, and (right) CMS~\cite{Chatrchyan:2013bka}, with all candidates weighted by their probability to be signal.
}
\label{fig:Bsmumu}
\end{figure}

Perhaps the most powerful of all Standard Model tests with rare \B decays is the search for $\Bds \to \mumu$ decays.
The decay rate is suppressed by all three of the features mentioned above, and in particular the helicity suppression due to the $V-A$ structure of the weak interaction is expected to be alleviated in models with extended Higgs sectors, such as supersymmetry.
Prior to LHC data taking, experimental limits~\cite{Abazov:2010fs,Aaltonen:2011fi} still allowed the $\Bs\to\mumu$ rate to be enhanced by more than a factor of ten compared to its Standard Model prediction of $(3.35 \pm 0.28) \times 10^{-9}$. 
However, a series of results from LHCb~\cite{LHCb-PAPER-2011-004,LHCb-PAPER-2011-025,LHCb-PAPER-2012-007}, CMS~\cite{Chatrchyan:2012rga} and ATLAS~\cite{Aad:2012pn} progressively reduced the phase space for new physics signals.
In late 2012, LHCb announced the first evidence for the $\Bs\to\mumu$ decay~\cite{LHCb-PAPER-2012-043}, and this was subsequently confirmed by both LHCb~\cite{LHCb-PAPER-2013-046} and CMS~\cite{Chatrchyan:2013bka} with their full Run~I data sets.
The signals, shown in Fig.~\ref{fig:Bsmumu}, when combined now reach the $5\sigma$ threshold for observation~\cite{LHCb-CONF-2013-012}.
The combined result for the branching fraction
$$
{\cal B}(\Bs \to \mumu) = (2.9 \pm 0.7) \times 10^{-9}
$$
is consistent with the Standard Model prediction.
Future updates with more statistics are nonetheless of great interest to reduce the uncertainty, to search for the $\Bd \to \mumu$ decay and to test the Standard Model prediction for the effective lifetime of the $\Bs\to\mumu$ decay.

New physics can also be searched for using $\Bz \to \Kstarz\mumu$ decays, or similar $b \to s\mumu$ transitions.
Such decays offer a wealth of asymmetries and angular observables that can be studied as functions of the dimuon invariant mass squared, $q^2$, several of which have been shown to have reduced theoretical uncertainty.
For example, although the absolute values of the branching fractions are hard to predict precisely, isospin and \CP asymmetries, between \Bp and \Bz or \B and \Bbar decays respectively, provide powerful null tests of the Standard Model.
Evidence of isospin asymmetry in $\B \to \kaon \mumu$ decays~\cite{LHCb-PAPER-2012-011} has not been confirmed in an update with larger statistics~\cite{LHCb-PAPER-2014-006}, and all \CP asymmetry measurements in $\B \to K^{(*)}\mumu$ decays are also consistent with zero~\cite{LHCb-PAPER-2012-021,LHCb-PAPER-2013-043}.

\begin{figure}[!t]
\centering
\includegraphics[width=0.50\textwidth]{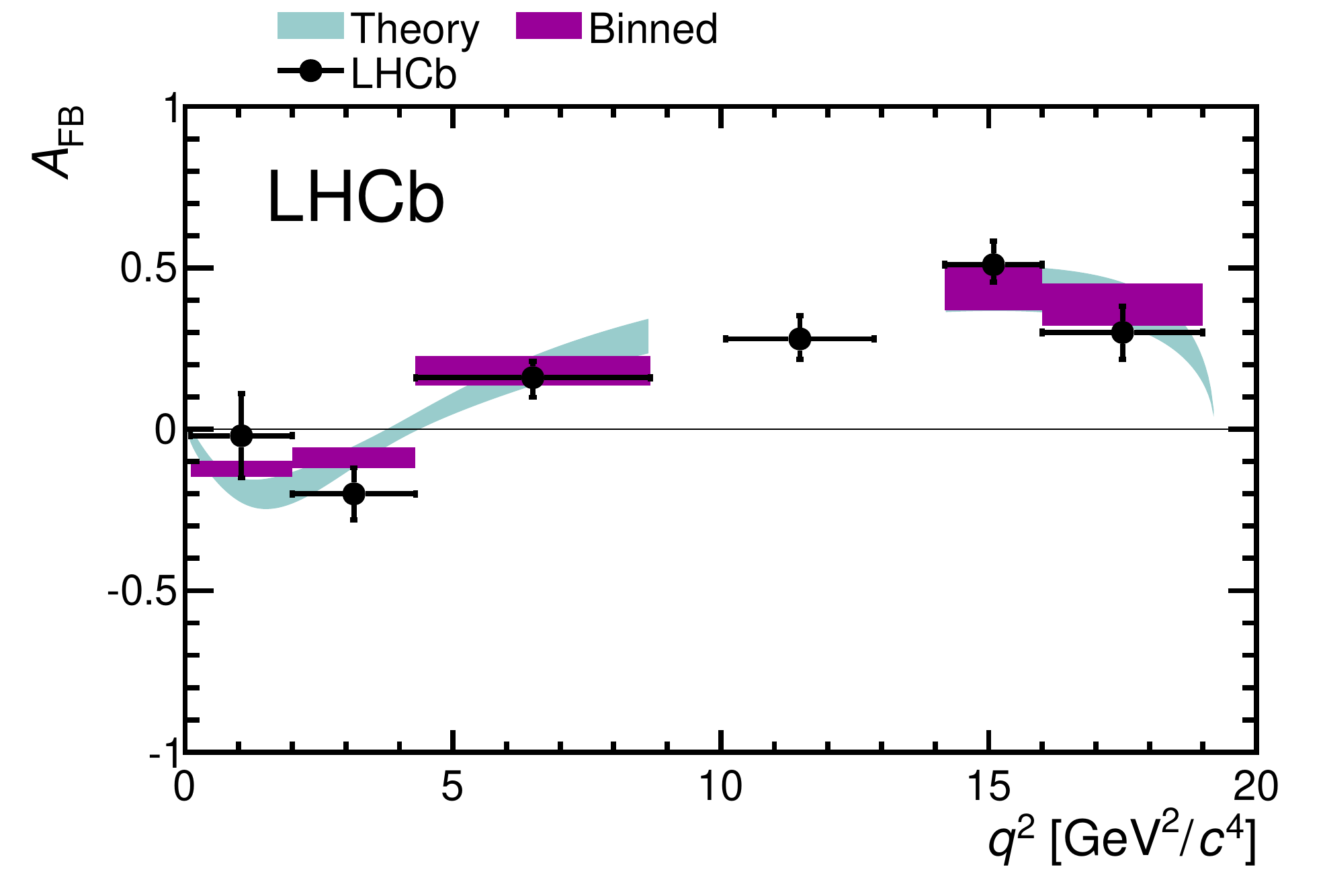}
\includegraphics[width=0.46\textwidth]{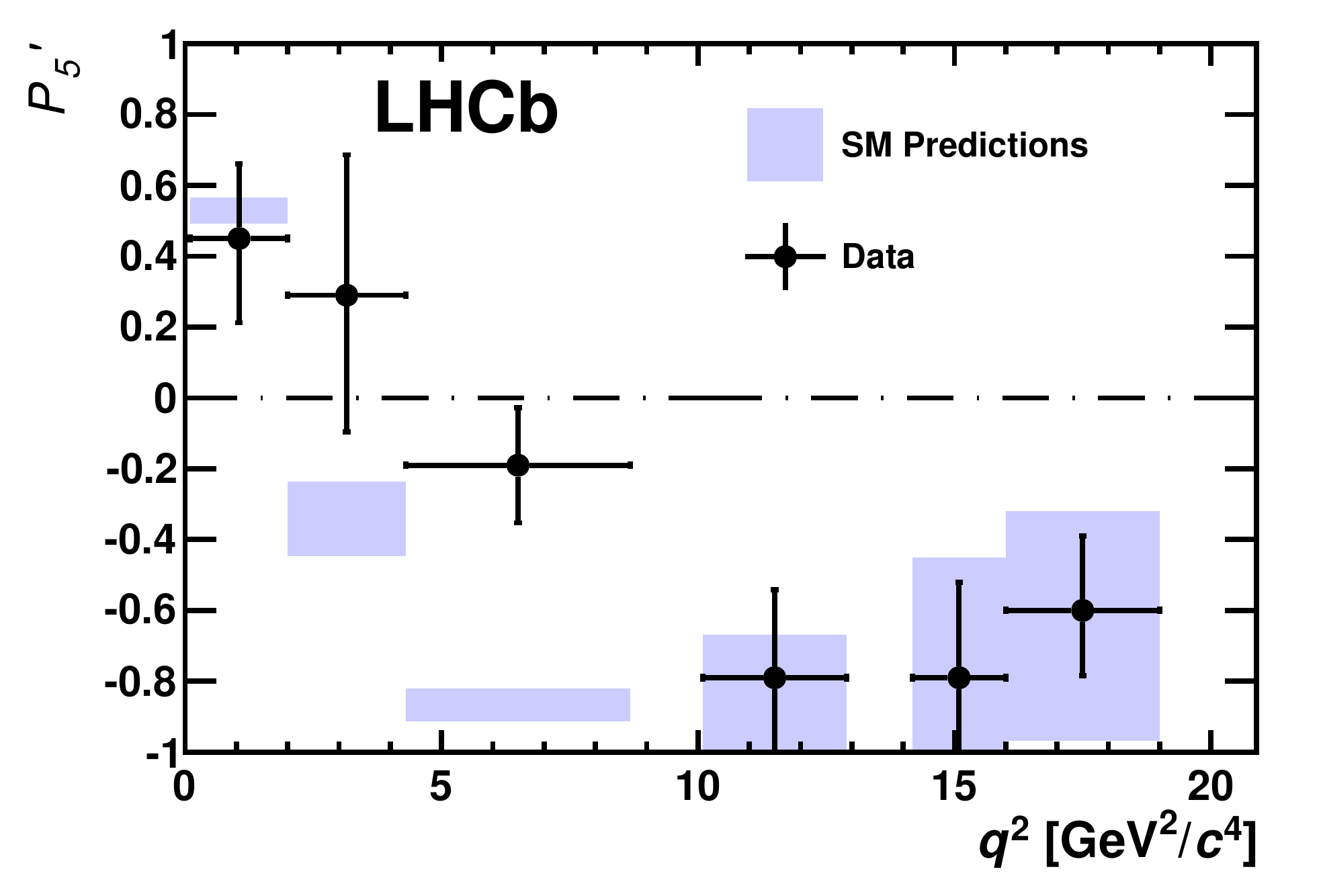}
\caption{\small
  Measurements of the (left) forward-backward asymmetry~\cite{LHCb-PAPER-2013-019} and (right) the observable $P_5^\prime$~\cite{LHCb-PAPER-2013-037} from LHCb.
}
\label{fig:Kstarmumu}
\end{figure}

Among other angular observables, the forward-backward asymmetry in $\Bz \to \Kstarz\mumu$ decays, \ie\ the difference in the average directions of the \mup and \mun particles in the rest frame of the decay, has received considerable interest.
In the Standard Model such an asymmetry arises, and varies with $q^2$ in a predictable way, due to interference between diagrams where the dimuon system is produced by virtual $\gamma$ and $Z^0$ bosons.
While previous measurements had shown evidence for a net forward-backward asymmetry when integrated over $q^2$~\cite{Aubert:2006vb,Wei:2009zv,Aaltonen:2011ja}, a measurement from LHCb~\cite{LHCb-PAPER-2013-019}, shown in Fig.~\ref{fig:Kstarmumu}, was the first to pin down the shape of the variation and measure the $q^2$ at which the asymmetry crosses zero.
LHCb has additionally extended the analysis to study further angular observables that have reduced theoretical uncertainty~\cite{LHCb-PAPER-2013-037}.
In one of them, labelled $P_5^\prime$, an interesting discrepancy with the Standard Model prediction is seen.
This result is based on the 2011 data sample, and an update with the full Run~I statistics together with improved theoretical predictions will help to determine whether the discrepancy is robust.
Further results on this topic are also anticipated from CMS~\cite{Chatrchyan:2013cda} and ATLAS~\cite{ATLAS-CONF-2013-038}.
As more data is collected similar observables can also be studied in $\Bs \to\phi\mumu$~\cite{LHCb-PAPER-2013-017} and $\Lb \to \Lz\mumu$~\cite{LHCb-PAPER-2013-025} decays as well as in $b \to d\mumu$ transitions such as $\Bp \to \pip\mumu$~\cite{LHCb-PAPER-2012-020}.

The study of the lowest region of the $q^2$ spectrum is of particular interest since the strong polarisation of the emitted photon in $b \to s \gamma^{(*)}$ transitions that is predicted in the Standard Model can be probed.
The non-negligible muon mass implies that it is preferable to carry out such studies with either $b \to s \gamma$ or $b \to s \epem$ decays.
The trigger capability of LHCb has allowed it to measure branching fractions and \CP asymmetries of $\Bz \to \Kstarz\gamma$ and $\Bs \to \phi \gamma$ decays~\cite{LHCb-PAPER-2011-042,LHCb-PAPER-2012-019}.  
The latter channel is particularly interesting because with more data the \CP violating parameters of its decay time distribution can be used to probe the photon polarisation with low theoretical uncertainty.
Angular distributions in $\Bz\to \Kstarz\epem$~\cite{LHCb-PAPER-2013-005} and $\Bp \to \Kp\pip\pim\gamma$~\cite{LHCb-PAPER-2014-001} decays provide complementary ways to test this Standard Model prediction, with the latter mode giving the first observation of non-zero photon polarisation.

\section{Summary and outlook}
\label{sec:summary}

The large data samples collected from high-energy $pp$ collisions during Run~I of the Large Hadron Collider have enabled dramatic progress in heavy flavour physics.
Significant advances have been achieved in understanding spectroscopy, \CP violation and rare decays.
The results are consistent with Standard Model predictions, although several puzzles and hints of discrepancies demand further investigation with larger data samples.
Since many of the results to date are based on data collected in 2011 only, further progress can be anticipated in the near future as the analyses are updated to the full Run~1 samples.

Substantial further improvement in precision will be possible with the data from the LHC Run~II.
The increased centre-of-mass energy of the collisions is expected to result in higher production cross-sections for heavy-flavoured particles, and therefore more useful events can in principle be recorded per $\invfb$ of data collected.
The actual gain will be determined by the trigger algorithms used for online event selection.

Beyond Run~II, the LHCb detector will be upgraded~\cite{LHCb-TDR-012} to allow improved trigger efficiencies with higher luminosity operation.  
With the LHCb upgrade, and the ongoing capability of ATLAS and CMS in the field of heavy flavour physics, this topic will remain a priority throughout the high luminosity LHC era.

\section*{Acknowledgements}

This work was supported by the Science and Technology Facilities Council (UK) and the European Research Council under FP7.
The authors are grateful to Sascha Turczyk for pointing out an error in a draft of the manuscript.

\section*{References}

\bibliography{LHCb-PAPER,references,quarkonia}

\end{document}

%% file: lhcb-symbols-def.tex
\usepackage{upgreek}

\def\MagUp {\mbox{\em Mag\kern -0.05em Up}\xspace}

\ifthenelse{\boolean{uprightparticles}}%
{

 \def\Pmu         {\ensuremath{\upmu}\xspace}

 \def\Ppi         {\ensuremath{\uppi}\xspace}

 \def\Ppsi        {\ensuremath{\uppsi}\xspace}

 \def\PDelta      {\ensuremath{\Delta}\xspace}                 
 \def\PXi      {\ensuremath{\Xi}\xspace}                 
 \def\PLambda      {\ensuremath{\Lambda}\xspace}                 
 \def\PSigma      {\ensuremath{\Sigma}\xspace}                 
 \def\POmega      {\ensuremath{\Omega}\xspace}                 
 \def\PUpsilon      {\ensuremath{\Upsilon}\xspace}                 
 
 \def\PB      {\ensuremath{\mathrm{B}}\xspace}                 
                  
 \def\PD      {\ensuremath{\mathrm{D}}\xspace}

 \def\PJ      {\ensuremath{\mathrm{J}}\xspace}                 
 \def\PK      {\ensuremath{\mathrm{K}}\xspace}

 \def\Pb      {\ensuremath{\mathrm{b}}\xspace}                 
 \def\Pc      {\ensuremath{\mathrm{c}}\xspace}                 
 \def\Pd      {\ensuremath{\mathrm{d}}\xspace}                 
 \def\Pe      {\ensuremath{\mathrm{e}}\xspace}

 \def\Pi      {\ensuremath{\mathrm{i}}\xspace}

 \def\Pp      {\ensuremath{\mathrm{p}}\xspace}

 \def\Ps      {\ensuremath{\mathrm{s}}\xspace}                 
                  
 \def\Pu      {\ensuremath{\mathrm{u}}\xspace}

}
{

 \def\Pmu         {\ensuremath{\mu}\xspace}

 \def\Ppi         {\ensuremath{\pi}\xspace}

 \def\Ppsi        {\ensuremath{\psi}\xspace}                 
                  
 \mathchardef\PDelta="7101
 \mathchardef\PXi="7104
 \mathchardef\PLambda="7103
 \mathchardef\PSigma="7106
 \mathchardef\POmega="710A
 \mathchardef\PUpsilon="7107
                  
 \def\PB      {\ensuremath{B}\xspace}                 
                  
 \def\PD      {\ensuremath{D}\xspace}

 \def\PJ      {\ensuremath{J}\xspace}                 
 \def\PK      {\ensuremath{K}\xspace}

 \def\Pb      {\ensuremath{b}\xspace}                 
 \def\Pc      {\ensuremath{c}\xspace}                 
 \def\Pd      {\ensuremath{d}\xspace}                 
 \def\Pe      {\ensuremath{e}\xspace}

 \def\Pi      {\ensuremath{i}\xspace}

 \def\Pp      {\ensuremath{p}\xspace}

 \def\Ps      {\ensuremath{s}\xspace}                 
                  
 \def\Pu      {\ensuremath{u}\xspace}

}

\makeatletter
\ifcase \@ptsize \relax
  \newcommand{\miniscule}{\@setfontsize\miniscule{4}{5}}
\or
  \newcommand{\miniscule}{\@setfontsize\miniscule{5}{6}}
\or
  \newcommand{\miniscule}{\@setfontsize\miniscule{5}{6}}
\fi
\makeatother

\DeclareRobustCommand{\optbar}[1]{\shortstack{{\miniscule (\rule[.5ex]{1.25em}{.18mm})}
  \\ [-.7ex] $#1$}}

\def\epem       {{\ensuremath{\Pe^+\Pe^-}}\xspace}

\def\mup        {{\ensuremath{\Pmu^+}}\xspace}
\def\mun        {{\ensuremath{\Pmu^-}}\xspace} 
\def\mumu       {{\ensuremath{\Pmu^+\Pmu^-}}\xspace}

\def\uquark    {{\ensuremath{\Pu}}\xspace}

\def\dquark    {{\ensuremath{\Pd}}\xspace}
\def\dquarkbar {{\ensuremath{\overline \dquark}}\xspace}

\def\squark    {{\ensuremath{\Ps}}\xspace}

\def\cquark    {{\ensuremath{\Pc}}\xspace}
\def\cquarkbar {{\ensuremath{\overline \cquark}}\xspace}

\def\bquark    {{\ensuremath{\Pb}}\xspace}
\def\bquarkbar {{\ensuremath{\overline \bquark}}\xspace}

\def\pion   {{\ensuremath{\Ppi}}\xspace}

\def\pip    {{\ensuremath{\pion^+}}\xspace}
\def\pim    {{\ensuremath{\pion^-}}\xspace}
\def\pipm   {{\ensuremath{\pion^\pm}}\xspace}
\def\pimp   {{\ensuremath{\pion^\mp}}\xspace}

\def\kaon    {{\ensuremath{\PK}}\xspace}
  \def\Kbar    {{\kern 0.2em\overline{\kern -0.2em \PK}{}}\xspace}

\def\KorKbar    {\kern 0.18em\optbar{\kern -0.18em K}{}\xspace}
\def\Kz      {{\ensuremath{\kaon^0}}\xspace}

\def\Kp      {{\ensuremath{\kaon^+}}\xspace}
\def\Km      {{\ensuremath{\kaon^-}}\xspace}
\def\Kpm     {{\ensuremath{\kaon^\pm}}\xspace}
\def\Kmp     {{\ensuremath{\kaon^\mp}}\xspace}
\def\KS      {{\ensuremath{\kaon^0_{\rm\scriptscriptstyle S}}}\xspace}

\def\Kstarz  {{\ensuremath{\kaon^{*0}}}\xspace}
\def\Kstarzb {{\ensuremath{\Kbar{}^{*0}}}\xspace}

  \def\Dbar    {{\kern 0.2em\overline{\kern -0.2em \PD}{}}\xspace}
\def\D       {{\ensuremath{\PD}}\xspace}

\def\DorDbar    {\kern 0.18em\optbar{\kern -0.18em D}{}\xspace}
\def\Dz      {{\ensuremath{\D^0}}\xspace}
\def\Dzb     {{\ensuremath{\Dbar{}^0}}\xspace}
\def\Dp      {{\ensuremath{\D^+}}\xspace}

\def\Dsp     {{\ensuremath{\D^+_\squark}}\xspace}
\def\Dsm     {{\ensuremath{\D^-_\squark}}\xspace}

\def\Dsmp    {{\ensuremath{\D^{\mp}_\squark}}\xspace}

\def\B       {{\ensuremath{\PB}}\xspace}
\def\Bbar    {{\ensuremath{\kern 0.18em\overline{\kern -0.18em \PB}{}}}\xspace}

\def\BorBbar    {\kern 0.18em\optbar{\kern -0.18em B}{}\xspace}
\def\Bz      {{\ensuremath{\B^0}}\xspace}
\def\Bzb     {{\ensuremath{\Bbar{}^0}}\xspace}
\def\Bu      {{\ensuremath{\B^+}}\xspace}

\def\Bp      {{\ensuremath{\Bu}}\xspace}

\def\Bd      {{\ensuremath{\B^0}}\xspace}
\def\Bs      {{\ensuremath{\B^0_\squark}}\xspace}
\def\Bds     {{\ensuremath{\B^0_{(\squark)}}}\xspace}
\def\Bsb     {{\ensuremath{\Bbar{}^0_\squark}}\xspace}

\def\Bc      {{\ensuremath{\B_\cquark^+}}\xspace}

\def\jpsi     {{\ensuremath{{\PJ\mskip -3mu/\mskip -2mu\Ppsi\mskip 2mu}}}\xspace}

  \def\Y#1S{\ensuremath{\PUpsilon{(#1S)}}\xspace}

\def\proton      {{\ensuremath{\Pp}}\xspace}

\def\Xires       {{\ensuremath{\PXi}}\xspace}

\def\Lz          {{\ensuremath{\PLambda}}\xspace}
\def\Lbar        {{\ensuremath{\kern 0.1em\overline{\kern -0.1em\PLambda}}}\xspace}
\def\LorLbar    {\kern 0.18em\optbar{\kern -0.18em \PLambda}{}\xspace}

\def\Lb      {{\ensuremath{\Lz^0_\bquark}}\xspace}

\def\to                 {\ensuremath{\rightarrow}\xspace}

\def\CP                {{\ensuremath{C\!P}}\xspace}

\newcommand{\dm}{{\ensuremath{\Delta m}}\xspace}
\newcommand{\dms}{{\ensuremath{\Delta m_{\squark}}}\xspace}

\newcommand{\DG}{{\ensuremath{\Delta\Gamma}}\xspace}
\newcommand{\DGs}{{\ensuremath{\Delta\Gamma_{\squark}}}\xspace}

\def\AT#1     {\ensuremath{A_{\mathrm{T}}^{#1}}\xspace}           

\def\C#1      {\ensuremath{\mathcal{C}_{#1}}\xspace}                       
\def\Cp#1     {\ensuremath{\mathcal{C}_{#1}^{'}}\xspace}                    
\def\Ceff#1   {\ensuremath{\mathcal{C}_{#1}^{\mathrm{(eff)}}}\xspace}        
\def\Cpeff#1  {\ensuremath{\mathcal{C}_{#1}^{'\mathrm{(eff)}}}\xspace}       
\def\Ope#1    {\ensuremath{\mathcal{O}_{#1}}\xspace}                       
\def\Opep#1   {\ensuremath{\mathcal{O}_{#1}^{'}}\xspace}                    

\newcommand{\tev}{\ifthenelse{\boolean{inbibliography}}{\ensuremath{~T\kern -0.05em eV}\xspace}{\ensuremath{\mathrm{\,Te\kern -0.1em V}}}\xspace}
\newcommand{\gev}{\ensuremath{\mathrm{\,Ge\kern -0.1em V}}\xspace}
\newcommand{\mev}{\ensuremath{\mathrm{\,Me\kern -0.1em V}}\xspace}
\newcommand{\kev}{\ensuremath{\mathrm{\,ke\kern -0.1em V}}\xspace}
\newcommand{\ev}{\ensuremath{\mathrm{\,e\kern -0.1em V}}\xspace}
\newcommand{\gevc}{\ensuremath{{\mathrm{\,Ge\kern -0.1em V\!/}c}}\xspace}
\newcommand{\mevc}{\ensuremath{{\mathrm{\,Me\kern -0.1em V\!/}c}}\xspace}
\newcommand{\gevcc}{\ensuremath{{\mathrm{\,Ge\kern -0.1em V\!/}c^2}}\xspace}
\newcommand{\gevgevcccc}{\ensuremath{{\mathrm{\,Ge\kern -0.1em V^2\!/}c^4}}\xspace}
\newcommand{\mevcc}{\ensuremath{{\mathrm{\,Me\kern -0.1em V\!/}c^2}}\xspace}

\def\mbarn{\ensuremath{\rm \,mb}\xspace}
\def\mub{\ensuremath{{\rm \,\upmu b}}\xspace}

\def\invfb   {\ensuremath{\mbox{\,fb}^{-1}}\xspace}

\def\ps   {\ensuremath{{\rm \,ps}}\xspace}
\def\fs   {\ensuremath{\rm \,fs}\xspace}

\def\invps{\ensuremath{{\rm \,ps^{-1}}}\xspace}

\newcommand{\stat}{\ensuremath{\mathrm{\,(stat)}}\xspace}
\newcommand{\syst}{\ensuremath{\mathrm{\,(syst)}}\xspace}

\def\gsim{{~\raise.15em\hbox{$>$}\kern-.85em
          \lower.35em\hbox{$\sim$}~}\xspace}
\def\lsim{{~\raise.15em\hbox{$<$}\kern-.85em
          \lower.35em\hbox{$\sim$}~}\xspace}

\def\pt         {\mbox{$p_{\rm T}$}\xspace}

\def\tell1  {TELL1\xspace}
\def\ukl1   {UKL1\xspace}

\newcommand{\eg}{\mbox{\itshape e.g.}\xspace}
\newcommand{\ie}{\mbox{\itshape i.e.}\xspace}